# Long-run Consequences of Health Insurance Promotion When Mandates are Not Enforceable: Evidence from a Field Experiment in Ghana [*]

Patrick Opoku Asuming, Hyuncheol Bryant Kim, and Armand Sim

May 2019

**Abstract**

We study long-run selection and treatment effects of a health insurance subsidy in Ghana, where mandates are not enforceable. We randomly provide different levels of subsidy (1/3, 2/3, and full), with follow-up surveys seven months and three years after the initial intervention. We find that a one-time subsidy promotes and sustains insurance enrollment for all treatment groups, but long-run health care service utilization increases only for the partial subsidy groups. We find evidence that selection explains this pattern: those who were enrolled due to the subsidy, especially the partial subsidy, are more ill and have greater health care utilization.

Key words: health insurance; sustainability; selection; randomized experiments
JEL code: I1, O12

[*] Contact the corresponding author, Hyuncheol Bryant Kim, at hk788@cornell.edu; Asuming: University of Ghana Business School; Kim: Department of Policy Analysis and Management, Cornell University; Sim: Department of Applied Economics and Management, Cornell University. We thank Ama Baafra Abeberese, Douglas Almond, Diane Alexander, Jim Berry, John Cawley, Esteban Mendez Chacon, Pierre-Andre Chiappori, Giacomo De Giorgi, Supreet Kaur, Robert Kaestner, Don Kenkel, Daeho Kim, Michael Kremer, Wojciech Kopczuk, Leigh Linden, Corrine Low, Doug Miller, Sangyoon Park, Seollee Park, Cristian Pop-Eleches, Bernard Salanie, and seminar participants at Columbia University, Cornell University, Seoul National University, and the NEUDC. This research was supported by the Cornell Population Center and Social Enterprise Development Foundation, Ghana (SEND-Ghana)." Armand Sim gratefully acknowledges financial support from the Indonesia Education Endowment Fund. All errors are our own. A previous version of this paper was circulated under the title "Long-Run Consequences of Health Insurance Promotion: Evidence from a Field Experiment in Ghana."

## 1. Introduction

Many poor households in developing countries lack access to health insurance, and their poverty is exacerbated by health-related problems (Dercon, 2002). In the absence of insurance, households bear a high proportion of medical expenses in the form of out-of-pocket (OOP) payments, and face financial constraints that act as significant barriers to health care access (Gertler and Gruber, 2002; Xu et al., 2003; Wagstaff, 2007).[1] Many developing countries have increasingly been instituting social health insurance schemes (SHIs) to help mitigate the effects of adverse health shocks, especially for the poor (WHO, 2005, 2010).[2] However, even though SHIs are theoretically mandates and offer low sign-up costs and generous benefits to increase enrollment, take-up and retention rates remain very low in many countries (Fenny et al., 2016), especially among the poorest households (Acharya et al., 2013).[3]

Achieving universal coverage or a high enrollment rate is important in terms of risk pooling and the sustainability of social health insurance; however, it is often difficult for developing countries to successfully impose mandates, primarily due to administrative constraints. For example, if those who are more ill or with larger health care service utilization are selected into social health insurance, financial burden of the program will increase and become less sustainable. There have been various efforts to promote health insurance enrollment and health care utilization, but many recent studies find such efforts have limited impact.[4]

Even after successfully promoting health insurance enrollment in the short run, retention and sustainable improvements in health service utilization and health status remain a challenge. The long-run effects of an intervention have important implications for policy. For example, an increased retention rate may yield economic and health benefits when individuals engage in health services on a regular and timely basis, which may improve the sustainability of the health insurance program. Nevertheless, this topic remains relatively understudied.

---

[1] The World Health Organization finds that OOP payments as a proportion of private expenditure on health reach 77.6% in low-income countries (WHO, 2015).

[2] Recent examples of countries that have instituted SHIs include Ghana, Kenya, Nigeria, Tanzania, and Vietnam. Countries in the process of instituting SHIs include Cambodia, Laos, Malaysia, Zimbabwe, and South Africa. (Wagstaff, 2010).

[3] In a rural district of Northern Ghana, our study area, the annual fees and premiums of the SHI are about $5; the program covers almost 95% of disease conditions without deductibles or copayments. However, by the end of 2010, the total active membership reached only 34% of the total population (National Health Insurance Authority (NHIA), 2011).

[4] For example, Wagstaff et al. (2016) and Capuno et al. (2016) find subsidy and information do not successfully promote health insurance enrollment. Thornton et al. (2010) find subsidy increases short-term enrollment but does not increase health care service utilization.

A subsidy is one of the few successful types of interventions used to promote health insurance enrollment (e.g., Thornton et al., 2010). However, an important question emerges regarding the level of subsidy. Different levels of subsidy (price) may attract people with different characteristics, and this selection may affect health care service utilization and health outcomes among the insured.[5] The screening effect of subsidy level has been studied for a few health products and services, such as facility delivery (Grepin et al, 2019), malaria bed nets (Cohen and Dupas, 2010), and chlorine for water purification (Ashraf, Berry, and Shapiro, 2010), but has not been investigated for health insurance in a developing country setting.[6]

This study aims to fill this gap in the literature through a field experiment. Similar to Kremer and Miguel (2007) and Dupas (2014), we employ experimental variations in exposure to a health product and follow the behavioral response in the long run. We randomly selected communities for the subsidy intervention and randomized different levels of subsidy (one-third, two-thirds, and full subsidy) for the insurance premiums and fees for one-year's coverage at the household level. To measure the impact of these interventions, we conduct a baseline survey and two follow-up surveys, one at seven months and the second at three years after the initial intervention.

This experiment has three main objectives. First, we study whether a subsidy for premiums and fees promotes health insurance enrollment in the short run. Second, we study whether the level of subsidy affects health insurance enrollment, health care service utilization, and health status to shed light on the potential selection effect of the level of health insurance subsidy. Third, we study whether a one-time intervention could have sustainable impacts on health insurance enrollment, health care service utilization, and health outcomes in the long run.

Three sets of results emerge. First, we find a significant increase in short-run insurance take-up. Those receiving one-third, two-thirds, and a full subsidy were 39.3, 48.3, and 53.8 percentage points, respectively, more likely to enroll in health insurance in the short-run. Three years after the initial intervention, we still observe increased enrollment. Those who received one-third, two-thirds, and a full subsidy were 17.3, 14.0, and 18.9 percentage points, respectively, more likely to enroll in health insurance in the long run.

[5] In addition, as Dupas (2014) explains, the price level may affect the long-run adoption of health products through the "anchoring" mechanism, where a previously encountered price may act as anchor and affect people's valuation of a product regardless of its intrinsic value.

[6] The burgeoning literature in developed countries, especially in the United States, has studied this topic extensively (see, for example, Einav, Finkelstein, and Levin, 2010 for a comprehensive overview).

Second, we find evidence of selection. Those who enrolled due to our intervention (compliers) are more ill and have larger health expenditures than those who did not enroll regardless of intervention (never-takers). Among compliers, individuals in the partial subsidy group are particularly more ill and have larger health expenditures than those in the full subsidy group. This evidence suggests that having to pay positive amount of premium and fees induces individuals to engage in selective enrollment and maximize net expected benefits of having insurance.

In addition, we also find that selection patterns are more prominent in the long run. One possible explanation is that in the long run individuals have more time to learn about their health types as well as cost and benefits of health insurance. These hypotheses are supported by Hendren (2019) who shows that the ex-ante value of insurance before an individual has gained information on her health type is understated.

Third, we do not find evidence of improvement in health status despite the increase in health care expenditures in the long run, especially for the partial subsidy groups. A possible explanation is that people become more sensitive to symptoms and/or aware of their illness, which may lead to a reporting problem: misperception of symptoms of an illness. This could happen when people with coverage make frequent contacts with health facilities (Finkelstein, et al, 2012).

In summary, this study shows that a short-term subsidy intervention can successfully sustain health insurance enrollment where mandates are not enforceable. At the same time, we also observe a selection pattern on observable characteristics, especially in the partial subsidy groups, which negatively affects the financial sustainability of social health insurance.

Our study contributes to three strands of literature. First, our study contributes to the broad empirical literature on the effects of health insurance coverage on health outcomes, which has so far produced mixed evidence. Thornton et al. (2010), Fink et al. (2013), and King et al. (2009) do not find evidence that health insurance affects overall health care expenditures and health outcomes in Nicaragua, Burkina Faso, and Mexico, respectively.[7] However, Miller, Pinto, and

[7] Further, Thornton et al. (2010) finds substitution between the use of health clinics covered by health insurance and those that are not covered, but overall utilization did not increase. King et al. (2009) find a decrease in catastrophic expenditure, but overall changes in health care service utilization are negligible.

Vera-Hernandez (2013) find that health insurance coverage improves health care service utilization and health outcomes in Colombia. In addition, Gruber, Hendren, and Townsend (2014) investigate Thailand's health care reform and find that increased access to health care among the poor could decrease infant mortality. In terms of OOP expenses, some studies observe no or adverse effects of insurance on such expenses (e.g., Thornton et al., 2010; Fink et al., 2013), while others find the opposite (e.g., Galárraga et al., 2010). Last, the existing literature on health insurance in developing countries largely focuses on short-run health effects.[8] To our knowledge, this study is the first to examine the effects of insurance coverage on both short- and long-run health outcomes in a low-income setting.

Second, our study contributes to the literature on sustainability of health intervention programs. This study is, to our knowledge, among the first to document evidence of the long-run effects of interventions on insurance enrollment retention in a developing country. While the idea of promoting sustainability is attractive, it is difficult to achieve in practice. The challenges in promoting sustainable health insurance enrollment could be even greater because health care services in developing countries are generally of low quality and unreliable.[9] The few studies on this topic include those by Kremer and Miguel (2007) and Dupas (2014). In contrast to Kremer and Miguel (2007), who find limited evidence that a subsidy promotes long-run adoption of worm treatment, Dupas (2014) finds that a one-time subsidy may boost long-run adoption of malaria bed nets.

It is important to note, however, that the long-run effect of a one-time health insurance intervention is quite different from that of health product adoption such as worm treatment and malaria bed nets. Having health insurance does not necessarily result in improved health status. To be successful, health insurance enrollment should promote health care service utilization and prevent moral hazard behaviors. In addition, learning about the effects of other health products, such as deworming medicine, bed nets, and water disinfectants, could be less setting-specific than the case of health insurance, where the quality of health care services could vary considerably across settings.

[8] In the US setting, a RAND experiment reports insignificant effects of insurance coverage on average health outcomes but finds negative effects on health outcomes for the more vulnerable subgroups (Newhouse and the Insurance Experiment Group, 1993). Relatively recent studies find positive effects of exposure to public health insurance during childhood on various long-term health outcomes (Currie, Decker, and Lin, 2008; Miller and Wherry, 2016; Boudreaux, Golberstein, and McAlpine, 2016).

[9] See, for example, Banerjee, Deaton, and Duflo, (2004), Goldstein, et al. (2013), and Das, et al., (2016) for illustrations of low health care quality in developing countries. Alhassan et al. (2016) provides illustrations for Ghana.

Third, our study complements a growing body of work that explains the role of pricing in take-up and use of health products and services in developing countries. We study whether the characteristics of people who remain enrolled in health insurance in the long run vary by the level of subsidy. The effect of prices on utilization of health products and services has received considerable attention recently. While proponents of user fees argue that cost-sharing is necessary for the sustainability of health programs (World Bank, 1993; Easterly, 2006), there is a concern that even a small fee may prevent those most in need from purchasing the product. Recent studies aiming to test the existence of the screening effects of higher prices on health product utilization find mixed results. While Ashraf, Berry, and Shapiro (2010) find that high prices stimulate product use through a screening effect in chlorine for water sanitation, Cohen and Dupas (2010) find no effect of higher prices on the use of malaria bed nets.

The remainder of this paper is structured as follows. Section 2 outlines the research context. Section 3 describes the experimental design and data. Section 4 presents the empirical strategy and Section 5 presents the main results. Section 6 concludes the paper.

## 2. Institutional Background

### 2.1. National Health Insurance Scheme (NHIS) in Ghana

The National Health Insurance Scheme (NHIS) in Ghana was established by the National Health Insurance Act (Act 560) in 2003. It aims to improve access to and the quality of basic health care services for all citizens, especially the poor and vulnerable (Ministry of Health, 2004). The law mandates that every citizen enroll in at least one scheme. However, in practice, there are no penalties for those who do not enroll. Most of the 170 administrative districts of Ghana operate their own District Mutual Health Insurance Scheme (DMHIS) (Gajate-Garrido and Owusua, 2013). [10] Each DMHIS accepts and processes applications, collects premiums (and fees), provides

[10] There are three types of insurance schemes in Ghana: District Mutual Health Insurance Schemes (DMHIS), Private Mutual Health Insurance Schemes (PMHIS), and Private Commercial Insurance Schemes (PCHIS). The focus of this study is DMHIS, which represents 96 percent of insurance coverage (GSS, GHS and ICF, 2009). They are operated and subsidized by the government through the National Health Insurance Fund (NHIF). PMHIS are non-profit non-subsidized schemes run by NGOs, religious bodies and cooperative societies. PCHISs are for profit schemes that do not receive government subsidies.

membership identification cards, and processes claims from accredited facilities for reimbursement.

Annual means-tested premiums, which are charged to informal sector workers, range from $5 to $ 32. However, owing to the lack of information on household incomes, rural districts tend to charge the lowest premiums, while urban districts charge higher premiums. Indigents, pregnant women, children under 18 years, and the elderly over 70 years are exempt from premiums but not registration fees.[11] All members, except for indigents and pregnant women, are required to pay registration fees when they first register and when they renew. Those who do not renew their membership by the due date pay penalties when they eventually renew their memberships.

The benefits package of the NHIS, which is the same across DMHISs, is very generous, albeit new members wait for three months before they can enjoy the insurance benefits. As described in Table A1, the package covers: 1) full outpatient and inpatient (surgery and medical) treatments and services, 2) full payment for medications on the approved list, 3) payments for referrals on the approved list, and 4) all emergencies. The NHIA (2010) estimates that 95% of disease conditions that affect Ghanaians are covered by the scheme. Those who enroll do not pay deductibles or copayments for health care service utilization by law; however, according to the USAID (2016), health care providers often charge unauthorized fees that are inaccurately described as copayments.[12]

Despite the low premiums and generous benefits, enrollment in the NHIS remains low. By the end of 2010, the total active membership stood at 34% of the population of Ghana (NHIA, 2011). Enrollment is particularly low among the poorest. A 2008 nationwide survey found that only 29% of the individuals in the lowest wealth quintile were active members of the scheme compared to 64% of households in the highest quintile (National Development Planning Commission, 2009).

In addition to the lack of affordability, negative perceptions of the NHIS explain the low enrollment rate. For example, Alhassan et al. (2016) note that those enrolled in the NHIS generally perceive they are not receiving good-quality health care, for reasons such as long wait times and the poor attitudes of health staff towards patients. Additionally, Fenny et al. (2016) observe that perceived quality of service and socio-cultural factors such as trust, bad attitudes of health facility staff, and drug shortage contribute to low enrollment and retention rates in Ghana.

[11] The law defines an indigent as “a person who has no visible or adequate means of income or who has nobody to support him or her and by the means test.” Specifically, an indigent is a person who satisfies all of these criteria: i) unemployed and has no visible source of income, ii) does not have a fixed place of residence according to standards determined by the scheme, iii) does not live with a person who is employed and who has a fixed place of residence, and iv) does not have any identifiably consistent support from another person.

[12] http://www.africanstrategies4health.org/uploads/1/3/5/3/13538666/country_profile_-_ghana_-_us_letter.pdf

## 2.2. Setting

This study was conducted in Wa West, a poor and remote rural district in Northern Ghana (Figure A1). It covers an area of approximately 5,899 $km^2$ and had a population of about 81,000 in 2010. Settlement patterns are highly dispersed, with most residents living in hamlets of about 100-200 people. This high dispersion, coupled with the poor road network, makes traveling within the district difficult and expensive. The economy is largely agrarian, with over 90% of the population working as farmers. Estimates from the 2006 Ghana Living Standard Survey indicate that average annual per-capita income and health expenditure in a rural savannah locality like Wa West were about $252 and $26, respectively (Ghana Statistical Service, 2008).

In the study area, even though the Community-Based Health and Planning Services (CHPS) has increased accessibility to health care services,[13] there are only six public health centers and no tertiary health facility.[14] During the study period, the district had only 15 professional nurses and no medical doctor (Nang-Beifua, 2010). The district also has a high disease burden. The most common cause of outpatient visits in the region is malaria, which accounts for one third of outpatient visits. Other common causes of outpatient visits are acute respiratory-tract infections and skin diseases.

The Wa West DMIHS was introduced in January 2007. In 2011, it charged a uniform premium of $5.46 (GHC 8.20) for adults (18-69) and a processing fee of $2.67 (GHC 4) for first-time members and $0.60 (GHC 1) for renewals. Late renewals incur a fee of $1.30 (GHC 2) in addition to full premiums for all years for which membership was not renewed.[15] The baseline enrollment rate in 2011 for the study sample is 20%.

[13] CHPS are community health facilities that provide primary health care. They are located within rural communities with limited access to larger hospitals and are manned by nurses. Among the services offered are treatment of common ailments (malaria and diarrheal diseases) and maternal and child care services.
[14] About 75% of the communities in the study sample were within 6 km (3.73 mi) of a health facility.
[15] The exchange rate at the time of the study was $1 = GHC 1.5.

## 3. Research Design

In this section, we discuss the original research design, data collection, definition and construction of key variables, descriptive statistics as well as the balance test of baseline characteristics.

### 3.1. Interventions

We begin by discussing the original study (Asuming, 2013). Original study introduced three interventions to 4406 individuals of 629 household in 59 communities: a subsidy for the insurance premiums and fees (*Subsidy*), an information campaign on the national health insurance (*Campaign*), and an option for individuals to sign up in their community instead of traveling to the district capital (*Convenience*). Interventions were overlapping and randomized at the community level. In total, we had eight study groups: *Subsidy* only, *Campaign* only, *Convenience* only, *Subsidy* + *Campaign*, *Subsidy* + *Convenience*, *Campaign* + *Convenience*, *Subsidy* + *Campaign* + *Convenience*, and a control group. Figure B.1 summarizes our original research design. The original study intended to analyze single intervention effects as well as complementarity among interventions, but it does not provide enough power to test original hypothesis.

We extend the original study by implementing a long-term follow-up survey and focus only on the *Subsidy* intervention, which is the most credible intervention in the original design.[16] The *Subsidy* intervention provides insurance premiums and fees to households in randomly selected communities. The level of subsidy was further randomized at the household level: one-third ($2.67), two-thirds ($5.40), or full ($8.13) subsidy (see Figure 1). Subsidies were given in the form of vouchers, which were distributed between November 2011 and January 2012, valid for two-month, and redeemable at the Wa West DMHIS center.[17] The voucher specified the names, ages, and genders of all household members, expiration date, and place of redemption. Households

[16] The underlying assumption to estimate unbiased causal effects of the *Subsidy* intervention is that there is no complementarity between *Subsidy* and other treatments, which we demonstrate in Table B.1. None of the eight complementarity tests (e.g., *Sub* + *Camp* = *Sub* & *Camp*) reject the null hypothesis of no complementarity at 5 % level. To further investigate the cleaner effects of subsidy variation, we restricted the sample to the control and *Subsidy* only groups (i.e., excluding *Subsidy* + *Campaign*, *Subsidy* + *Convenience*, and *Subsidy* + *Campaign* + *Convenience*). We provide the estimation results in Tables B.2 (effects on enrollment), B.3 and B.4 (short- and long-run effects on health care utilization), and B.5 and B.6 (short- and long-run effects on health outcomes). In general, the results are similar to the main results. There are some differences, but they do not affect the interpretation of our main findings.

[17] The voucher could also be used to either initiate or renew insurance membership. Those who did not enroll at the baseline (80 %) could use the voucher anytime. Those who had already enrolled at the baseline (20 %) could only use the voucher if their existing renewal was due within the voucher's validity period. Otherwise they could not use the voucher.

that did not receive a full subsidy were informed about the extra amount needed to register all members.[18]

### 3.2. Data Collection

The study sample includes 2,954 individuals from 418 households in 44 communities. We conducted the baseline survey in September 2011 and implemented the intervention in October 2011. Two follow-up surveys were conducted, one at seven months and the second three years after the intervention. The baseline survey collected information on demographic characteristics, employment, health status, health care service utilization, enrollment in the NHIS, and health behaviors for all household members.

The first follow-up survey collected information on health care service utilization, health status, and health behaviors. In the second follow-up survey, we collect sets of information similar to those in the first follow-up survey but with greater detail to improve the quality of the data. For example, we asked for specific dates and the respondent's status since the first follow-up for up to three episodes of several important illnesses, such as malaria, acute respiratory diseases, and skin diseases. As a result, there are some differences in the construction of short- and long-run utilization measures that prevent a direct comparison of health care service utilization and health status in these survey periods.[19]

The main outcome variables of interest are health insurance enrollment, health care service utilization, health status, and health behaviors.[20] Health care service utilization is measured by health facility visits in the last four weeks and last six months as well as OOP health expenditure.

[18] For one-third or two-thirds subsidy households, vouchers took one of two forms: specified and unspecified. If a household received a specified subsidy voucher, its members were listed on the voucher, along with the specific amount of subsidy for each of them. Thus, reallocation of a subsidy within a household was not possible. If a household received an unspecified subsidy voucher, reallocation of the subsidy was possible because the voucher only showed the total amount of subsidy for the whole household, not the specific amount for each member.

[19] The health facility visit variable in the first follow-up survey is constructed from the following question: "The last time (in the last four weeks/last six months) (NAME) was ill or injured, did he/she visit any health facility?" However, in the second follow-up survey, the same variable is constructed from questions about respondents' visits during illness episodes. For example, an individual is said to visit a health facility in the last six months if his/her illness episode occurred in the last six months and he/she sought treatment in the health facility. This different structure of questions suggests that the magnitude of effects between the short- and long-run are not directly comparable. Studies have documented the role of recall periods on self-reported health status and health care utilization in developing countries (e.g., Das et al, 2012).

[20] Health insurance enrollment is measured at the individual level. Self-rated health status, which is restricted to those aged 18 years or older, is only available in the follow-up surveys. Health behaviors are measured for those aged 12 years or older.

Health status is measured by the number of days of illness in the last four weeks, the indicator and the number of days an individual was unable to perform normal daily activities due to illness as well as self-rated health status.[21] Health behavior is measured based on whether the respondent was sleeping under bed nets and using safe water technologies.[22]

The attrition rate in the first follow-up survey was relatively low (5 %) but increases in the second follow-up survey (21 %), as shown in Table A2.[23] The short- and long-run attrition rates are not systematically correlated with our interventions.

### 3.3. Baseline Characteristics and Balance Test

Table 1 presents the summary statistics of baseline characteristics and balance tests between the intervention and control groups. Panels A, B, and C report the average values of the individual, household, and community characteristics. Columns 1 and 2 report the average characteristics for all respondents and control groups, respectively. The average respondent is about 24 years old and 48% are male. About 20% were enrolled in the NHIS at the baseline survey, and 36% had ever registered with the scheme. In terms of health characteristics, 12% reported a sickness or injury in the last four weeks, about 4% visited a health facility in the last month, and 14% made a positive OOP health expenditure. The average household lives within 5.4 km of a health facility and 20 km from the district capital.

Our empirical approach requires a balance of baseline characteristics between the intervention and control groups that could affect outcome variables. To test this assumption, we compare the means of the variables at the baseline (Table 1). Columns 3 to 5 present results from regressions of each variable on control and subsidy level indicators. Column 6, which reports the *p*-values from the equality test, shows that only 2 out of 31 tests are statistically significant at the 10% level. We also compare the baseline differences between each subsidy level group in Columns 8 to 10. 5 out of 93 *t*-tests are statistically significant at the 10% level. Overall, these results suggest that our randomization is successful in creating balance across the control and treatment groups.

---

[21] Variables for limited normal daily activities were derived from the following questions: "During the four weeks, did (NAME) have to stop his/her usual activities because of this (illness/injury)?" and "For how many days (in the last one month) was (NAME) unable to do his/her usual activities?"

[22] We ask only about sleeping under a bed net in the baseline and short-run follow-up surveys, but we ask for more details on bed net and safe water technology use in the long-run survey.

[23] The main reasons for attrition in the first follow-up survey are deceased (17%), traveled (61%), relocated to other districts (16%), and others (6%). Information on reasons for attrition is not available in the second follow-up survey.

## 4. Estimation Framework

To measure the effects of our intervention on various outcomes, we estimate the following reduced-form intent-to-treat (ITT) effect of each level of subsidy:

$$y_{ihc} = \gamma_0 + \gamma_1 1/3Subsidy_{ihc} + \gamma_2 2/3Subsidy_{ihc} + \gamma_3 FullSubsidy_{ihc} + \theta X_{ihc} + \delta Z_{hc} + \omega V_c + \epsilon_{ihc} \quad (1)$$

where $y_{ihc}$ denotes the outcomes for individual *i* of household *h* in community *c*. The outcomes of interest include NHIS enrollment, health care service utilization, health status, and health behaviors. **X** denotes a vector of baseline individual covariates, such as indicator variables for age, gender, religion, ethnicity, and schooling. Household covariates **Z** include household size and a wealth index indicator (poor third, middle third, and rich third).[24] Community covariates **V** include distance to the nearest health facility and to the NHIS registration center. [25] We also control for a baseline measure of the dependent variable to improve precision. The results are robust when we exclude the baseline controls (results not shown). Estimations employ a linear probability model. For each outcome, we present its short- and long-run estimations.

We cluster standard errors at the community level[26] to account for possible correlation in the error terms within the same community.[27] We also perform 1,000 draws of a wild-cluster bootstrap percentile *t*-procedure suggested by Cameron et al (2008) to address concerns about small number of clusters, which could lead to downward-biased standard errors (Bertrand et al., 2004; Cameron et al., 2008).[28]

To obtain the effects of insurance coverage for compliers, we conduct a two-stage least squares (2SLS) regression, where the first-stage regression equation is Equation (1) with health insurance enrollment in the short run as the dependent variable. We estimate the following second-stage regression:

$$y_{ihc} = \alpha_0 + \alpha_1 \widehat{Enrolled}_{ihc} + \theta X_{ihc} + \delta Z_{hc} + \omega V_c + \epsilon_{ihc} \quad (2)$$

where we instrument for short-run enrollment. Then, we capture the local average treatment effect for those who were induced to enroll in health insurance as results of our subsidy intervention. [29]

---

[24] The wealth index is obtained through a principal components analysis with dwelling characteristics (e.g., number of rooms and bedrooms in the house), enterprise (e.g., ownership of any private non-farm enterprise), livestock (e.g., number of chickens and pigs), and other assets (e.g., motorcycles and bicycles).
[25] In addition, we controlled for indicators for *Subsidy + Campaign, Subsidy + Convenience,* and *Subsidy + Campaign + Convenience.*
[26] To account for correlation within household, we also cluster standard errors at the household level. The results do not change our main conclusion (results available upon request).
[27] Individuals in the same community are not completely independent of each other, especially in terms of health-related outcomes (Srinivasan et al., 2003). There are also studies that find important roles of communities on a host of socio-economic outcomes (Kling et al., 2007; Chetty et al., 2016).
[28] Our study has 44 clusters. While there is no clear threshold for too few clusters, the number could vary between 20 and 50 (Cameron and Miller, 2015).

Because we estimate Equation (1) for many different outcome variables in health care utilization and health status domains, a multiple hypothesis testing problem may occur. The probability we incorrectly reject at least one null hypothesis is larger than the conventional significance level. We address this concern using two methods. First, we group outcome variables into a domain and take the average standardized treatment effect in each domain, as suggested by Kling, Liebman, and Katz (2007) and Finkelstein et al. (2012). For the health care utilization domain, we group five outcome measures including intensive and extensive measures of health facility visits in the last four weeks and last six months and OOP expense incidence. For the health status domain, we group four outcomes including self-rated health status, number of days of illness, inability to perform normal activities, and the number of days lost to illness. Second, we apply the free step-down resampling procedure to adjust the family-wise error rate, that is, the probability of incorrectly rejecting one or more null hypotheses within a family of hypotheses (Westfall and Young, 1993). Family-wise adjusted *p*-values of each family are obtained from 10,000 simulations of estimations.[30]

## 5. Results

### 5.1. Impacts on Insurance Take-up, Sustainability, and Price Elasticity

Figure 2 shows the enrollment rates of the control and treatment groups at the baseline, short-run follow-up, and long-run follow-up surveys by level of subsidy. In general, it shows that enrollment rate increases with subsidy level in the short and long run, but the impacts attenuate over time. We observe the largest incremental increase in enrollment rate between receiving zero (control group) and one-third subsidy in the short run, but smaller incremental increases in the subsequent levels of subsidy. In the long run, the treatment group is still more likely to enroll in health insurance, but the differences among the one-third, two-thirds, and full subsidy groups become insignificant.

[29] We assume that income effect of the subsidy (\$2.7 - \$8.1) is small and negligible given that average income of the households in catchment area is \$252.

[30] These two methods serve different objectives. The first method is relevant for drawing general conclusions about the treatment effects on health care utilization and health status. The second method is more appropriate for examining the treatment effect of a specific outcome belonging to a set of tests.

Table 2 presents the formal regression results. We present robust standard errors in parentheses as well as two-tailed wild cluster bootstrap *p*-values in square brackets. Our results show that the effects on enrollment attenuate but are sustained over time. Column 1 of Panel A shows that overall subsidy intervention increases short-run insurance enrollment by 43.6 percentage points (160%). Long-run enrollment also increases by 20.6 percentage points (90%) (Column 2).

In terms of the level of subsidy, receiving a one-third, two-thirds, and full subsidy is associated with, respectively, a 39.3, 48.3, and 53.8 percentage points higher likelihood of enrolling in insurance in the short run than the control group (Column 1). Even though the enrollment rate of the one-third subsidy group is lower than that of the two-thirds and full subsidy groups in the short run, the enrollment rate of the one-third subsidy group is at least as large as those of the two-thirds and full subsidy groups in the long run (*p-value* > 0.6).

The short-run arc elasticities are large. Overall, when price decreases from $8.13 to $0, demand for health insurance increases from 27.2% to 81.0% (arc elasticity is -0.54).[31] The estimated arc elasticity is close to the elasticity of preventive health products in developing countries, such as -0.6 for chlorine, a disinfectant that prevents water-borne diseases in Zambia (Ashraf, Berry, and Shapiro, 2010), and -0.37 for insecticide-treated bed nets for malaria prevention in Kenya (Cohen and Dupas, 2010). The estimated arc elasticity is also similar to that of preventive health products in developed countries, such as -0.17 and -0.43 for preventive health care in the United States (Newhouse and the Insurance Experiment Group, 1993) and -0.47 for cancer screening in Korea (Kim and Lee, 2017).

[31] Arc elasticity estimates were obtained using the following formula: $[(Y_a - Y_b)/(Y_a + Y_b)]/[(P_a - P_b)/(P_a + P_b)]$, where $Y$ and $P$ denote enrollment rate and price, respectively. The short-run arc elasticity estimates when price increases from $0 to $2.67, $2.67 to $5.40, and $5.40 to $8.13 are 0.04, 0.19, and 2.10, respectively. Comparing the arc elasticity in a zero-price setting to those in other settings could be problematic because the denominator, $(P_a - P_b)/(P_a + P_b)$, is always 1 if $P_b = 0$. Moreover, people tend to treat a zero price not only as a decrease in cost but also as an extra benefit (Shampanier, Mazar, and Ariely, 2007). These results must be interpreted with this caveat.

Our finding that a larger subsidy may lead to higher health insurance enrollment corresponds to Finkelstein et al. (2017). However, our finding is contradictory to the special zero price argument suggesting that individuals act as if pricing a good as free not only decreases its cost but also adds to its benefit (Shampanier et al., 2007). For example, several studies find a larger decrease between zero and small non-zero prices in demand for malaria bed nets (Dupas, 2014) and HIV testing (Thornton, 2008). In contrast, we find a very large incremental increase in enrollment between zero and the one-third subsidy (full and two-thirds price) but no significant difference between the two-thirds and full subsidy (one-third and zero price). A possible explanation for this finding is the framing of the price of health insurance. Unlike Thornton (2008) and Dupas (2014), our subsidy intervention focuses on the level of subsidy instead of the level of price, and, therefore, the largest response to the intervention is found between zero and a small (one-third) subsidy.

### 5.2. Selection into Health Insurance

Selection into social health insurance could have important implications for the financial sustainability of the program, especially when mandates are not enforceable, in that people who are more ill or those with larger health care service utilization could be more likely to select into the program.

We first show evidence of selective retention in health insurance by level of health care service utilization. Those who have larger health care service utilization are more likely to remain enrolled in health insurance, as shown by the standardized treatment effects (Panels A and B of Table A3).

To gain further insight on selection into health insurance, we compare the individual and household characteristics of compliers, always-takers, and never-takers. The impacts we estimate are driven by compliers who enroll in health insurance due to our subsidy intervention. Following Almond and Doyle (2011) and Kim and Lee (2017), we calculate the mean characteristics and test the differences among compliers, always-takers, and never-takers.

To do so, we first define a binary variable $T$, an indicator for whether an individual is assigned to the treatment group (*Subsidy*). Next, we define a binary variable $H$, an indicator for whether an individual is enrolled in health insurance. Lastly, we define $H_T$ as the value $H$ would have if $T$ were either 0 or 1. Hence, $E(X|H_1 = 1)$ presents the mean value characteristics of treated individuals who enrolled in health insurance. Under the assumption of existence of the first stage, monotonicity, and independence, $E(X|H_1 = 1)$ can be written as:

$$E(X|H_1 = 1) = E(X|H_1 = 1, H_0 = 1)\cdot P(H_0 = 1|H_1 = 1) + E(X|H_1 = 1, H_0 = 0)\cdot P(H_0 = 0|H_1 = 1) \quad (3)$$

Equation (3) implies that $E(X|H_1 = 1)$ is a sum of always-takers and compliers components. $E(X|H_1 = 1, H_0 = 0)$ represents the characteristics of compliers. $E(X|H_1 = 1, H_0 = 1) = E(X|H_0 = 1)$ holds from the monotonicity assumption. $P(H_0 = 1)$, the proportion of always-takers, and $P(H_1 = 0)$, the proportion of never-takers, can be directly measured from the sample. $P(H_0 = 1)$, the proportion of always-takers can be thus measured by $P_a$, the proportion of insurance takers in the control group. Similarly, the proportion of never-takers, $P(H_1 = 0)$, can also be measured by $P_b$, the proportion of insurance non-takers in the treatment group. The proportion of compliers is $1- P_a - P_b$. Therefore, $P(H_0 = 1|H_1 = 1)$ and $P(H_0 = 0|H_1 = 1)$ are $\frac{P_a}{P_c+P_a}$ and $\frac{P_c}{P_c+P_a}$, respectively.[32]

By rearranging equation (3), the characteristic of compliers can be calculated as follows:

$$E(X|H_1 = 1, H_0 = 0) = \frac{P_c+P_a}{P_c} \times [E(X|H_1 = 1) - \frac{P_a}{P_c+P_a} \times E(X|H_0 = 1)] \quad (4)$$

Table 3 presents the summary statistics of the entire sample, compliers, always-takers, and the never-takers for short-run selection (Columns 1 to 4) and long-run selection (Columns 8 to 11). Columns 5 to 7 report the *t*-statistics for the mean comparison between compliers and always-takers, compliers and never-takers, and always-takers and never-takers in the short run. Columns 12 to 14 report similar statistics in the long run. By comparing compliers and never-takers, we find that our subsidy intervention attracted people who were more ill and had larger health expenditure, especially in the long run (Column 13). For example, compliers were more likely to have limited daily activities in the last four weeks compared to never-takers, and the differences became larger and more significant in the long run.

Next, we explore the selection pattern by level of subsidy by comparing compliers of the full subsidy and partial subsidy intervention. To do so, we restrict the sample to those who were insured in the *Subsidy* treatment group. The reasonable assumption that we impose is that always-takers in the full and partial subsidy groups are the same. Since we restrict our sample to those insured in the treatment group, which consists of compliers and always-takers, any difference between full and partial subsidy groups in the restricted sample is due to the compositional changes of compliers. Table 4 presents the results of 24 regressions where we regress each health characteristic on an indicator of full subsidy. The last two rows in Panels A and B report the average standardized effects for health status and health care utilization in the short and long run, respectively. The results show that partial subsidy compliers are more likely to be ill and have larger health care expenditure in the long run but not in the short run.

[32] The estimated share of compliers, always-takers, and never-takers are 47.4%, 27.1%, and 25.5% in the short run, and 24.3%, 23.0%, and 52.7% in the long run, respectively.

In summary, we find that, in general, compliers are more ill and have larger health expenditures than never-takers. Among the compliers, those in the partial subsidy group are more ill and have larger health expenditure than those in the full subsidy group. In addition, Tables 3 and 4 show that the selection patterns are more prominent in the long run, suggesting heterogeneous impacts of interventions on health care utilization by level of subsidy, especially in the long run.

Stronger selection in the partial subsidy group compared to the full subsidy group is not surprising. Those in the partial subsidy group need to pay positive amount of insurance premiums and fees, compared to zero-cost for the full-subsidy group. Those with partial subsidy may enroll in health insurance only if they expect the net gain to be positive, that is, expected benefits of health insurance are greater than the cost. Stronger selection in the long run could be because health insurance is an experience good, a service where product characteristics are easier to observe upon consumption (Nelson, 1970). For example, those with health insurance can afford to make more frequent contacts with medical services. They can collect more private health information and learn more about costs and benefits of health insurance than those without insurance. These hypotheses are supported by Hendren (2019) who shows that the ex-ante value of insurance before an individual has gained information on her health type is understated.

### 5.3. Impacts on Health Care Services Utilization

Table 5 presents the effects on utilization of health care services in the short run (Columns 1 to 6). Column 6 presents average standardized effects; Panels A and B present ITT and 2SLS results, respectively. We report bootstrap and family-wise *p*-values in square and curly brackets, respectively. The long-run effects are presented in Table 6.

We find that insurance coverage leads to an increase in utilization of health care services in both the short and long runs, which corresponds to the fact that health insurance enrollment is sustained in the long run (Panel A1 of Tables 5 and 6). It is worth noting that an increase in health care service utilization in the long run is at least as high as that in the short run (Columns 6 of Tables 5 and 6, respectively) even though the enrollment rate decreased.

The results regarding health care utilization by level of subsidy are interesting (Panels A2 and A3 of Tables 5 and 6). Even though the increase in long-run health insurance enrollment is similar across subsidy levels and the short-run increase in health care utilization is greater in the full subsidy group, we find evidence of an increase in health care utilization in the long run only for the partial subsidy group, not the full subsidy group. This result suggests that selection could be important for explaining the increase in health care utilization through health insurance promotion.

We also study the impacts on OOP expenses (Column 5). We find limited evidence that health insurance prevents OOP expenses either in the short or long run.[33] There are a few possible explanations for this finding. First, as we described earlier, most services are free under the NHIS, but health care providers often charge unauthorized fees as copayments. Second, medicine is often in short supply at the public health centers, and those who receive a diagnosis may purchase medicine from a private pharmacy. Third, those without health insurance often use traditional or herbal medicine which is inexpensive, and therefore, substitution from traditional medicine to formal health care does not decrease OOP expenses.

---

[33] Again, the size effects in the short- and long-run are not directly comparable because the short- and long-run OOP expenses are constructed differently. In the short run, respondents were asked about more general OOP expenses, but in the long run, OOP expenses only included those related to the treatment of several important illnesses (e.g., malaria, skin diseases, and acute respiratory infection). Specifically, for the short-run OOP expense, we use the individual's response to the following question: "On (NAME's) most recent visit to a health facility, did he/she pay any money from his/her own pocket at a health facility in the last six months?" On the other hand, to construct the long-run OOP expense, we use information on whether individuals made positive OOP expenses in each illness episode (i.e., malaria, acute respiratory infection, and skin diseases) that occurred in the last six months.

### 5.4. Impacts on Health Status

Table 7 presents the effects on health status in the short run (Columns 1 to 5). Column 5 presents average standardized effects.[34] The long-run effects are presented in Table 8. Panel A1 of Table 7 shows that insurance coverage improves health status in the short run.[35] However, Panel A1 of Table 8 shows that short-run positive health effect seems to disappear in the long run even though health insurance enrollment and health care service utilization continue to increase, as shown in Tables 2, 5, and 6. Panel B, which shows the 2SLS results, confirms a similar pattern: the emergence of short-run positive health effects (although most are not statistically significant) dissipate in the long run. We even find negative health effects on the number of sick days and daily activities in the long run (Columns 2 to 4 of Table 8). The negative health consequences in the long run are mainly driven by those in the partial subsidy group (Columns 2 to 4 of Panels A2 and A3 of Table 8) who also experienced an increase in health care utilization.[36]

Negative health consequences despite increased health care utilization in the long run appears contradictory. One possible explanation is that people become more sensitive to symptoms and/or aware of their illness, which may lead to a reporting problem. This could happen when people make frequent contacts with health facilities (Finkelstein, et al, 2012). Those who experience health care services could learn about the specific symptoms of illnesses, and thus become more sensitive about their health status. Also, those who receive a diagnosis could be more aware of the times or periods they were sick. As a result, they are more likely to report being ill.

Unfortunately, we are not able to test these explanations with our data, especially because our health status measures are self-reported, and it is difficult to know whether the negative self-reported health status reflects an actual deterioration in physical health. More research is needed to verify more precise mechanisms through which health insurance enrollment and health care utilization may result in a decline in self-reported health status.

---

[34]We group outcome measures into two domains (i.e., health care utilization and health status) and then take the average standardized treatment effect in each domain, as suggested by Kling, Liebman, and Katz (2007) and Finkelstein et al (2012). Specifically, we stack individual outcome data within each domain and estimate a single regression equation for each domain to obtain the average standardized treatment effect. For the health care utilization domain, we group five outcome measures including intensive and extensive measures of health facility visits in the last four weeks and last six months and out-of-pocket expense incidence. For the health status domain, we group four outcomes including self-reported health status, number of days of illness, inability to perform normal activities, and the number of days lost to illness.

[35] Self-rated health status is measured only for those aged 18 years and above.

[36] To help shed light on the lack of long-run health outcomes, we investigate individuals' health behaviors regarding the use of malaria bed nets and safe water technologies. We find some suggestive evidence on the decrease in the overall health investments in the full subsidy group, which is not consistent with the results in health utilization and status (Column 5 of Table A4).

## 6. Conclusion

This study examines the long-term consequences of one-time short-run subsidy interventions on health insurance enrollment, health care service utilization, and health outcomes in the long run, especially when mandates are not enforceable. In addition, we study the role of pricing in health insurance by measuring important behavioral responses to different levels of subsidy (i.e., one-third, two-thirds, and full subsidy). In Northern Ghana, we implement three randomized subsidy interventions to promote health insurance enrollment. We then use the resulting variation in insurance coverage to estimate the effect of insurance coverage on utilization of health care services, OOP expenses, and health status and behaviors.

We highlight three main findings. First, our interventions significantly promoted enrollment in the short run, and while the impacts attenuate, the positive impacts remained three years after the initial intervention implementation. Specifically, those treated with one-third, two-thirds, and full subsidies were 39.3, 48.3, and 53.8 percentage points, respectively, more likely to enroll health insurance in the short run, and 17.3, 14.0, and 18.9 percentage points, respectively, more likely to enroll in the long run.

Second, we find evidence of selection, especially in the long run. Compliers are more ill and have larger health expenditures than never-takers, and this pattern is more prominent in the long run. Among compliers, individuals in the partial subsidy group are particularly more ill and have larger health expenditures than those in the full subsidy group. As a result, health care expenditures of the partial subsidy group, who more selectively enrolled in health insurance, increases in the long run, even though health insurance enrollment rates are similar across levels of subsidy. Third, we do not find evidence of improvement in self-reported health status despite the increase in health care utilization in the long run.

Critics of the Ghanaian NHIS have argued that the scheme is overly generous and financially unsustainable because of the huge percentage of NHIS members under premium exemption without co-payment (Alhassan et al., 2016). The financial state of the NHIS program can be further endangered by selection behaviors we find in this study.

Our results provide some policy implications for health insurance in other developing countries. As mentioned earlier, the general impression of health care services under Ghana's NHIS is not positive implying that the long-run take-up of health insurance and service utilization could be higher in a setting with better institutions and health care services. Nevertheless, policy makers should be cautious of the presence of selection and behavioral responses since they are often difficult to predict and, importantly, may endanger financial stability of an insurance program.

Taken together, these findings highlight that even though short-run interventions successfully increase health insurance enrollment, their long-run success in promoting health status could depend on behavioral responses such as selection. Our findings suggest that as health insurance continues to be introduced in developing counties, both careful enforcement of mandatory health insurance enrollment to prevent selection and establishment of policies to encourage desirable health behaviors need to be considered.

**References**

Acharya, A., Vellakkal, S., Taylor, F., Masset, E., Satija, A., Burke, M. and Ebrahim, S. 2013. "The Impact of Health Insurance Schemes for the Informal Sector in Low- and Middle-Income Countries: A systematic Review," *World Bank Research Observer.*

Alhassan RK, Nketiah-Amponsah E, Arhinful DK. 2016. "A Review of the National Health Insurance Scheme in Ghana: What Are the Sustainability Threats and Prospects?" *PLoS ONE* 11(11): 1-16.

Almond, D. and Doyle, J.J., 2011. After midnight: A regression discontinuity design in length of postpartum hospital stays. *American Economic Journal: Economic Policy, 3*(3), pp.1-34.

Ashraf, Nava, James Berry, and Jesse Shapiro. 2010. "Can Higher Prices Stimulate Product Use? Evidence from a Field Experiment in Zambia," *American Economic Review,* 100 (5): 2283-2413.

Asuming, Patrick O. "Getting the Poor to Enroll in Health Insurance, and its effects on their health: Evidence from a Field Experiment in Ghana." Job Market Paper–Columbia University (2013).

Banerjee, A., A. Deaton, and E. Duflo. 2004. “Health, Health Care, and Economic Development,” *American Economic Review,* 94: 326–330.

Boudreaux, M.H., E. Golberstein, and D. McAlpine. 2016. “The Long-Term Impacts of Medicaid Exposure in Early Childhood: Evidence from the Program’s Origin,” *Journal of Health Economics*, 45:161-175.

Cameron, A.C. and D. Miller. 2015. “A Practitioner’s Guide to Cluster-Robust Inference,” *Journal of Human Resources,* 50 (2): 317-372.

Capuno J.J., A.D. Kraft, S. Quimbo, C.R. Tan, Jr., and A. Wagstaff. 2016. “Effects of Price, Information, and Transaction Cost Interventions to Raise Voluntary Enrolment in a Social: Health Insurance Scheme: A Randomized Experiment in the Philippines,” *Health Economics*, 25(6): 650-662.

Chetty, R., N. Hendren, and L. Katz. 2016. “The Effects of Exposure to Better Neighborhoods on Children: New Evidence from the Moving to Opportunity Project.” *American Economic Review* 106 (4): 855–902.

Cohen, J. and P. Dupas. 2010. “Free Distribution or Cost-sharing? Evidence from a Randomized Malaria Prevention Experiment,” *Quarterly Journal of Economics,* 125 (1): 1-45.

Currie, J., S. Decker, and W. Lin. 2008. “Has Public Health Insurance for Older Children Reduced Disparities in Access to Care and Health Outcomes?” *Journal of Health Economics*, 27: 1567-1581.

Das, J., J. Hammer, and C. Sánchez-Paramo. 2012. “The Impact of Recall Periods on Reported Morbidity and Health Seeking Behavior,” *Journal of Development Economics* 98: 76–88.

Das, Jishnu, Alaka Holla, Aakash Mohpal, and Karthik Muralidharan. 2016. “Quality and Accountability in Health Care Delivery: Audit-Study Evidence from Primary Care in India,” *American Economic Review*, 106: 3765–3799.

Dercon, Stefan. 2002. “Income Risk, Coping Strategies, and Safety Nets,” *The World Bank Research Observer*, 17: 141–166.

Dupas, P. 2014. “Short-run Subsidies and Long-run Adoption of New Health Products: Evidence from a Field Experiment,” *Econometrica*, 82 (1): 197-228.

Easterly, William. 2006. *The White Man’s Burden: Why the West’s Effort to Aid the Rest Have Done So Much Ill and So Little Good*. New York: Penguin Press.

Einav, Liran, Amy Finkelstein, and Jonathan Levin. 2010. “Beyond Testing: Empirical Models of Insurance Markets,” *Annual Review of Economics,* 2: 311-336.

Fink, F., P. Robyn, A. Sié, Rainer Sauerborn. 2013. ”Does Health Insurance Improve Health? Evidence from a Randomized Community-Based Insurance Rollout in Rural Burkina Faso,” *Journal of Health Economics,* 32:1043-1056.

Finkelstein, A., N. Hendren, and M. Shepard. 2017. “Subsidizing Health Insurance for Low-Income Adults: Evidence From Massachusetts,” *NBER Working Paper* 23668.

Finkelstein, Amy, S. Taubman, B. Wright, M. Bernstein, J. Gruber, J. P. Newhouse, H. Allen, K. Baicker, and Oregon Health Study Group. 2012. “The Oregon Health Insurance Experiment: Evidence from the First Year,” *Quarterly Journal of Economics,* 127 (3): 1057-1106.

Fenny, Ama, Anthony Kusi, Daniel Arhinful, and Felix Asante. 2016. “Factors contributing to low uptake and renewal of health insurance: a qualitative study in Ghana,” *Global Health Research and Policy*: 1-10.

Gajate-Garrido, G. and Rebecca Owusua. 2013. "The National Health Insurance Scheme in Ghana: Implementation Challenges and Proposed Solutions," *IFPRI Discussion Paper No.01309.*

Galarraga, Omar, Sandra G. Sosa-Rubi, and Sergio Sesma-Vazquez. 2010. "Health insurance for the poor: impact on catastrophic and out-of-pocket health expenditures in Mexico," *European Journal Health Economics*, 11:437–447.

Gertler, P. and Gruber, J. 2002. "Insuring Consumption Against Illness," *American Economic Review*, 92(1): 51-70.

Ghana Statistical Service. 2008. Ghana Living Standards Survey: Report of the Fifth Round (GLSS 5), Accra, Ghana: Ghana Statistical Service.

Ghana Statistical Service (GSS), Ghana Health Service (GHS), and ICF Macro. 2009. Ghana Demographic and Health Survey 2008. Accra, Ghana: GSS, GHS, and ICF Macro.

Goldstein, Markus, Joshua Zivin, J. Habyarimana, Cristian Pop-Eleches, and H. Thirumurthy. 2013. "The Effect of Absenteeism and Clinic Protocol on Health Outcomes: The Case of Mother-to-Child Transmission of HIV in Kenya," *American Economic Journal: Applied Economics,* 5(2): 58–85.

Grepin, Karen A., James Habyarimana, and William Jack. 2019. "Cash on Delivery: Results of a Randomized Experiment to Promote Maternal Health Care in Kenya." *Journal of Health Economics,* 65: 15-30.

Gruber, J., Hendren, N. and Townsend, R.M., 2014. The great equalizer: Health care access and infant mortality in Thailand. *American Economic Journal: Applied Economics,* 6(1): 91-107.

Hendren, N. 2019. "Measuring Ex-Ante Welfare in Insurance Markets" Unpublished manuscript.

Kim H. and Lee S. 2017. "When Public Health Intervention is not Successful: Cost Sharing, Crowd-Out, and Selection in Korea's National Cancer Screening Program," *Journal of Health Economics,* 53: 100-116.

King, G., Gakidou, E., Imai, K., Lakin, J., Moore, R.T., Nall, C., Ravishankar, N., Vargas, M., Tellez-Rojo, M.M., Avila, J.E., Avila, M.H. and Llamas, H.H. 2009. "Public Policy for the Poor? A Randomized Assessment of the Mexico Universal Health Insurance Programme," *The Lancet*, 373: 1447-1454.

Kling, J. R., J. B. Liebman, and L. F. Katz. 2007. "Experimental Analysis of Neighborhood Effects," *Econometrica,* 75 (1): 83–119.

Kremer, M. and Miguel, E. 2007. "The Illusion of Sustainability," *Quarterly Journal of Economics,* 122(3): 1007-1065.

Miller, Grant, Diana Pinto, and M. Vera-Hernandez. 2013. "Risk Protection, Service Use, and Health Outcomes under Colombia's Health Insurance Program for the Poor," *American Economic Journal: Applied Economics*, 5 (4): 61-91.

Miller, Sarah and Laura R. Wherry. 2016. "The Long-Term Effects of Early Life Medicaid Coverage," *Unpublished Working Paper*.

Ministry of Health. 2004. "Legislative Instrument on National Health Insurance. Accra: National Parliament of Ghana Press".

Nelson, Phillip. 1970. "Information and Consumer Behavior," *Journal of Political Economy* Vol. 78, No. 2 (Mar. - Apr., 1970), pp. 311-329

Newhouse, J.P. and the Insurance Experiment Group. 1993. *Free for All? Lessons from the RAND Health Insurance Experiment*. Cambridge, MA: Harvard University Press.

National Development Planning Commission. 2009. “2008 Citizen’s Assessment of the National Health Insurance Scheme,” Accra: National Development Planning Commission.

Nang-Beifua, A. 2010. Health Sector Half-Year Performance Report - Upper West Region, Accra, Ghana: Ghana Health Service.

National Health Insurance Authority. 2010. *National Health Insurance Scheme: Annual Report 2009,* Accra, Ghana: National Health Insurance Authority.

National Health Insurance Authority. 2011. *National Health Insurance Scheme: Annual Report 2010,* Accra, Ghana: National Health Insurance Authority.

Shampanier, K., Mazar, N and Ariely, D. 2007. “Zero as a Special Price? The True Value of Free Products,” *Marketing Science,* 26(6): 742-757.

Thornton, R.L., Hatt, L.E., Field, E.M., Islam, M., Diaz, F.S., and Gonzalez, M.A. 2010. “Social Security Health Insurance for the Informal Sector in Nicaragua: A Randomized Evaluation,” *Health Economics,* 19(S1):181-206.

Thornton, Rebecca. 2008 “The Demand for, and Impact of, Learning HIV Status,” *American Economic Review,* 98 (5): 1829–1863.

USAID, 2016. *Health Insurance Profile: Ghana.* Report presented at the Financial Protection and Improved Access to Health Care: Peer-to-Peer Learning Workshop held in Accra, Ghana.

Wagstaff, A. 2007. “The Economics Consequences of Health Shocks: Evidence from Vietnam,” *Journal of Health Economics* 26(1): 82-100.

Wagstaff, A. 2010. “Social Health Insurance Re-examined.” *Journal of Health Economics* 19: 503–517.

Wagstaff, A., H.T. Hong Nguyen, H. Dao and S. Bales. 2016. “Encouraging Health Insurance for the Informal Sector: A Cluster Randomized Experiment in Vietnam,” *Health Economics*, 25(6): 663-674.

Westfall, Peter H, and S Stanley Young. 1993. *Resampling-based multiple testing:Examples and methods for p-value adjustment.* Vol. 279, John Wiley & Sons.

WHO. 2005. “Sustainable Health Financing, Universal Coverage, and Social Health Insurance,” In: 58th World Health Assembly. Agenda Item 13.16 Edition. Geneva: WHO.

WHO. 2010. *The World Health Report: Health Systems Financing: the Path to Universal Coverage*. Geneva, Switzerland: World Health Organization Press.

WHO. *World health statistics,* 2015. Geneva: World Health Organization.

World Bank. 1993. *World Development Report: Investing in Health*. Oxford; New York; Toronto and Melbourne: Oxford University Press.

Xu, K., D. Evans, K. Kawabata, R. Zeramdini, J. Klavus, and C. Murray. 2003. “Household Catastrophic Health Expenditure: A Multi-Country Analysis,” *The Lancet* 362 (9378): 111-7.

**Figures and Tables**

Figure 1: Study Design

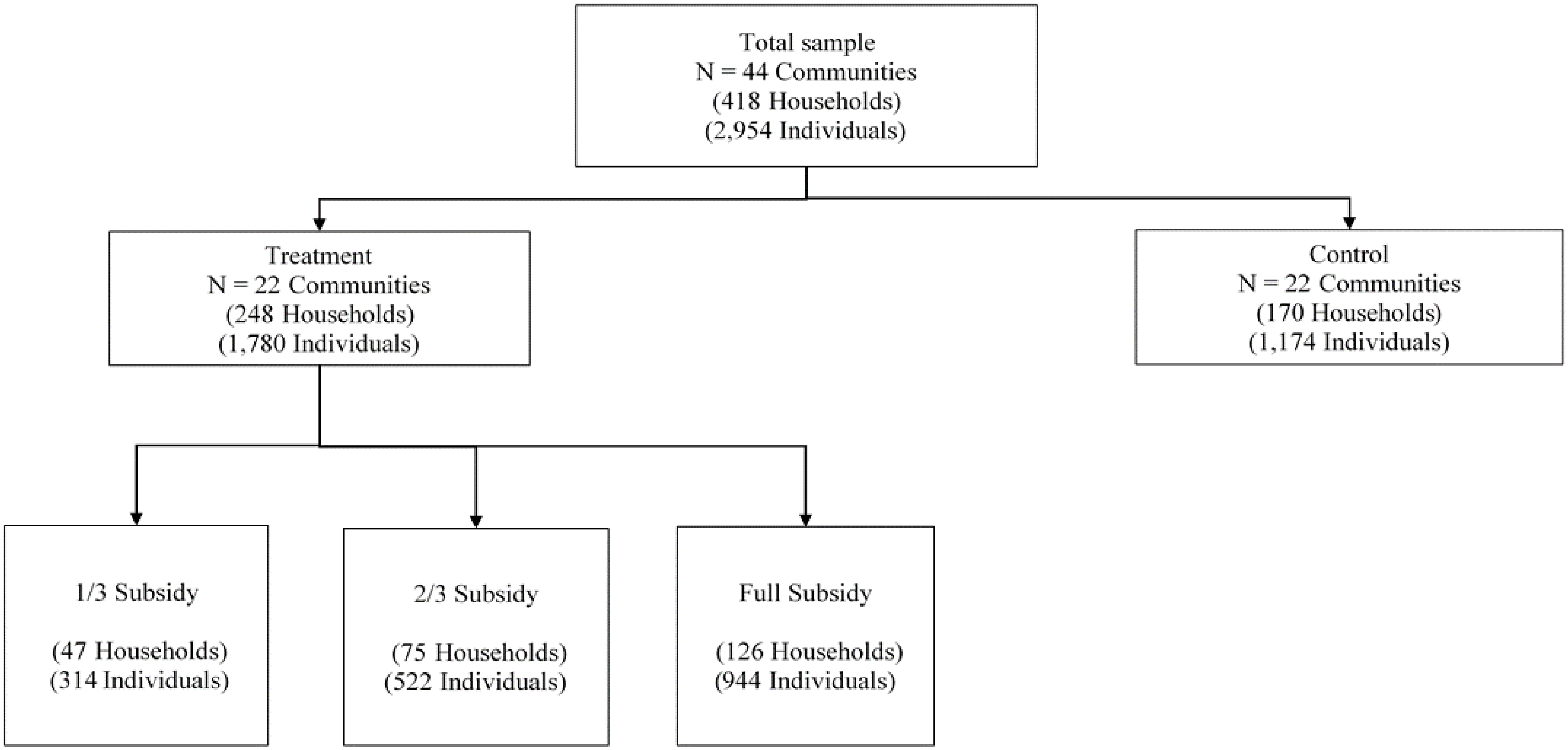

Figure 2: Enrollment Rate by Subsidy Level at Baseline, Short Run, and Long Run

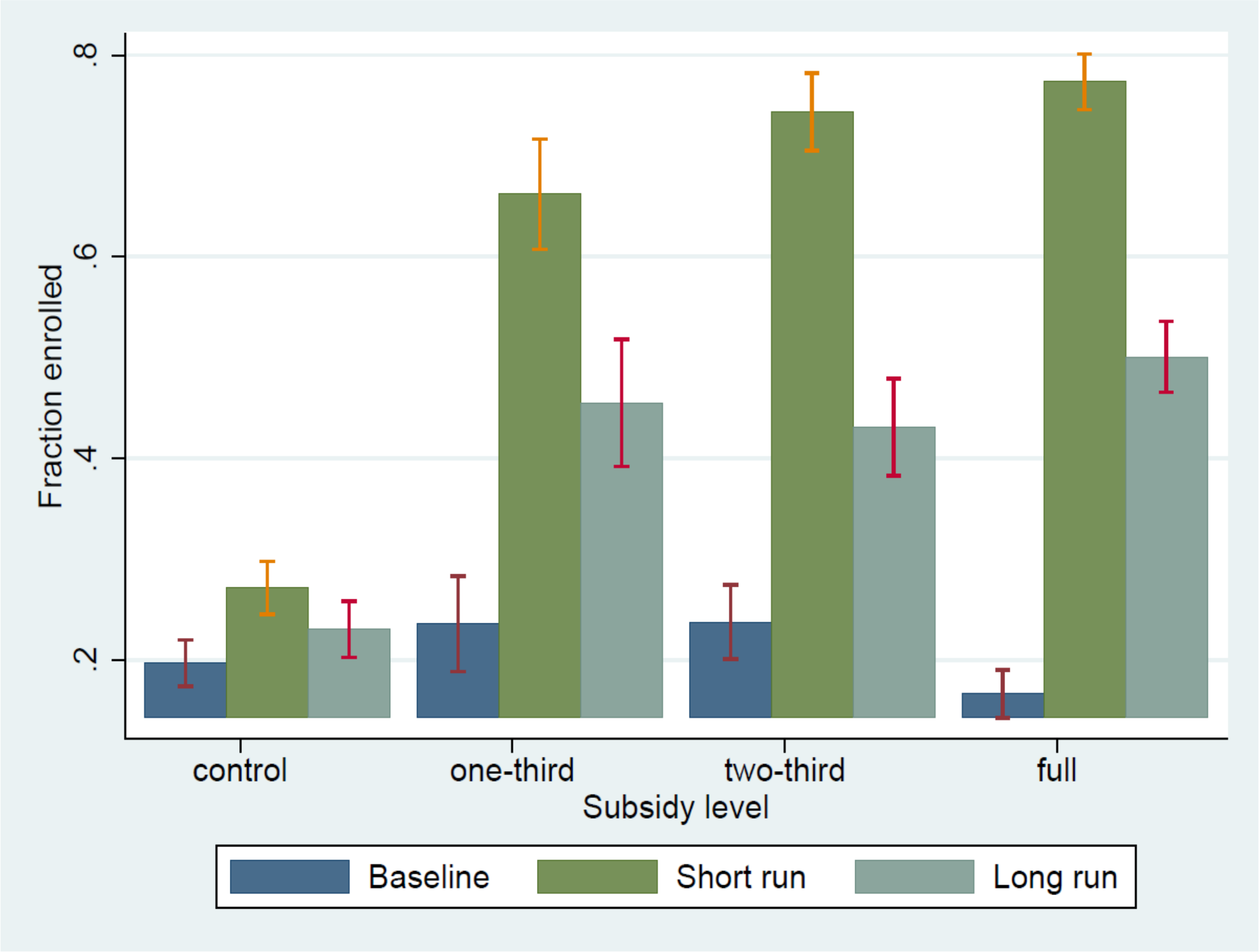

Notes: This figure shows means of enrollment rates of each subsidy-level group at baseline, short run, and long run. Sample includes those who received subsidy and the control group. The vertical lines indicate 95% confidence intervals.

## Table 1: Baseline Characteristics by Subsidy Level Group

| Variable | Mean | | Difference between subsidy level and control | | | | N | Difference between each subsidy level | | |
|---|---|---|---|---|---|---|---|---|---|---|
| | Full | Control | One-third | Two-thirds | Full | p-value | | One-Third vs Two-Thirds | One-Third vs Full | Two-Thirds vs Full |
| | (1) | (2) | (3) | (4) | (5) | (6) | (7) | (8) | (9) | (10) |
| ***Panel A: Individual Characteristics*** | | | | | | | | | | |
| Age | 23.780 | 24.310 | 1.180 | -0.775 | -1.620 | 0.390 | 2,954 | 1.955 | 2.800 | 0.844 |
| Male | 0.481 | 0.475 | 0.009 | -0.010 | 0.022 | 0.578 | 2,954 | 0.019 | -0.013 | -0.031 |
| Christian | 0.417 | 0.373 | 0.073 | 0.102 | 0.058 | 0.801 | 2,954 | -0.029 | 0.015 | 0.044 |
| Dagaaba (ethnic group) | 0.517 | 0.458 | 0.153 | 0.208 | 0.017 | 0.370 | 2,954 | -0.055 | 0.136 | 0.191 |
| Has some formal education | 0.335 | 0.337 | -0.022 | -0.015 | 0.009 | 0.976 | 2,954 | -0.007 | -0.031 | -0.025 |
| Has a health condition (≥ 6 months) | 0.071 | 0.072 | -0.002 | -0.002 | -0.004 | 0.996 | 2,954 | -0.001 | 0.001 | 0.002 |
| Probably sick next year | 0.441 | 0.436 | 0.030 | 0.023 | -0.005 | 0.809 | 2,845 | 0.007 | 0.035 | 0.028 |
| Overall illness | | | | | | | | | | |
| Ill in the last month (1/0) | 0.123 | 0.105 | 0.039 | 0.048 | 0.016 | 0.532 | 2,954 | -0.010 | 0.023 | 0.032 |
| # of days ill in the last month | 0.918 | 0.846 | 0.505 | 0.208 | -0.056 | 0.565 | 2,927 | 0.296 | 0.560 | 0.264 |
| Could not do normal activities in the last month (1/0) | 0.076 | 0.060 | 0.011 | 0.039 | 0.023 | 0.428 | 2,919 | -0.028 | -0.012 | 0.016 |
| # of days could not perform normal activities in the last month | 0.544 | 0.480 | 0.134 | 0.138 | 0.079 | 0.867 | 2,815 | -0.004 | 0.055 | 0.060 |
| Malaria | | | | | | | | | | |
| Ill in the last month (1/0) | 0.046 | 0.041 | -0.006 | 0.028 | 0.004 | 0.483 | 2,931 | -0.034 | -0.010 | 0.024 |
| # of days ill in the last month | 0.243 | 0.220 | -0.049 | 0.182 | -0.011 | 0.721 | 2,909 | -0.231 | -0.037 | 0.194 |
| Could not do normal activities in the last month (1/0) | 0.025 | 0.018 | -0.002 | 0.023 | 0.011 | 0.431 | 2,919 | -0.025 | -0.013 | 0.012 |
| # of days could not perform normal activities in the last month | 0.146 | 0.128 | -0.036 | 0.056 | 0.036 | 0.693 | 2,815 | -0.092 | -0.072 | 0.020 |
| Visited health facility in the last month | 0.039 | 0.036 | 0.033 | 0.023 | -0.015 | 0.301 | 2,435 | 0.010 | 0.048 | 0.038 |
| Visited health facility in the last six months | 0.074 | 0.074 | 0.025 | 0.008 | -0.014 | 0.639 | 2,954 | 0.016 | 0.038 | 0.022 |
| Number of visits in the last month | 0.066 | 0.063 | 0.062 | 0.042 | -0.036 | 0.117 | 2,443 | 0.020 | 0.098 | 0.078* |
| Visited health facility in the last month for malaria treatment | 0.010 | 0.011 | -0.004 | 0.002 | -0.004 | 0.925 | 2,435 | -0.006 | -0.0005 | 0.006 |
| Made out of pocket expense in the last six months | 0.136 | 0.133 | -0.009 | 0.059 | -0.021 | 0.306 | 2,954 | -0.067 | 0.012 | 0.079* |
| Ever enrolled in NHIS | 0.358 | 0.302 | 0.179** | 0.084 | 0.071 | 0.090* | 2,954 | 0.096 | 0.108 | 0.012 |
| Currently enrolled in NHIS | 0.198 | 0.197 | 0.039 | 0.041 | -0.030 | 0.690 | 2,954 | -0.002 | 0.069 | 0.071 |
| Slept under mosquito nets (12 years old or older) | 0.501 | 0.448 | 0.192** | 0.140 | 0.025 | 0.111 | 1,720 | 0.053 | 0.168 | 0.115 |
| Use safe drinking water technology (12 years old or older) | 0.024 | 0.039 | -0.039 | -0.019 | -0.020 | 0.109 | 1,286 | -0.020 | -0.020 | 0.001 |
| ***Panel B: Household Characteristics*** | | | | | | | | | | |
| HH Size | 8.703 | 8.454 | -0.187 | 0.051 | 0.813 | 0.517 | 2,953 | -0.238 | -0.999 | -0.761 |
| Number of children under 18 | 5.141 | 4.952 | 0.054 | -0.125 | 0.641 | 0.578 | 2,954 | 0.179 | -0.587 | -0.766 |
| Owns farming land | 0.553 | 0.506 | 0.118 | -0.007 | 0.112 | 0.363 | 2,674 | 0.125 | 0.006 | -0.119 |
| Owns mosquito net | 0.711 | 0.690 | 0.020 | 0.146 | -0.032 | 0.139 | 2,477 | -0.125* | 0.052 | 0.177* |
| Household assets (principal component score) | 0.601 | 0.266 | 0.580 | 0.269 | 0.705** | 0.061* | 2,953 | 0.311 | -0.126 | -0.436 |
| ***Panel C: Community Characteristics*** | | | | | | | | | | |
| Distance to NHIS regist (km) | 20.010 | 20.370 | 4.347 | 3.447 | -4.466 | 0.303 | 2,954 | 0.900 | 8.812 | 7.912 |
| Distance to health facility (km) | 5.394 | 5.166 | 0.221 | -0.687 | 1.017 | 0.149 | 2,954 | 0.908 | -0.796 | -1.704** |

Notes: Columns 1 and 2 report mean of all respondents and control group. Columns 3 to 5 present results from regressions of each variable on control and subsidy level indicators (1/3, 2/3, and full). Column 6 reports the *p*-value from a joint test of equality of the three coefficients reported in Columns 3 to 5. Column 7 reports total number of observations. Columns 8 to 10 present results from separate regressions of each variable on one-third and two-thirds subsidy levels (Column 8), on one-third and full subsidy levels (Column 9), and on two-thirds and full subsidy levels (Column 10). Robust standard errors are clustered at community level. *, **, and *** denote statistical significance at 10 %, 5 %, and 1 %. levels, respectively.

Table 2: Effects of Subsidy on NHIS Enrollment

| | Enrollment | |
|---|---|---|
| | Short-run | Long-run |
| | (1) | (2) |
| **Panel A** | | |
| Any Subsidy | 0.436*** | 0.206*** |
| | (0.048) | (0.059) |
| | [0.000] | [0.007] |
| R-squared | 0.342 | 0.160 |
| **Panel B** | | |
| Partial subsidy | 0.444*** | 0.154* |
| | (0.054) | (0.079) |
| | [0.000] | [0.094] |
| Full subsidy | 0.530*** | 0.192* |
| | (0.060) | (0.097) |
| | [0.000] | [0.117] |
| R-squared | 0.351 | 0.183 |
| **Panel C** | | |
| 1/3 subsidy | 0.393*** | 0.173** |
| | (0.072) | (0.083) |
| | [0.001] | [0.080] |
| 2/3 subsidy | 0.483*** | 0.140 |
| | (0.060) | (0.086) |
| | [0.000] | [0.153] |
| Full subsidy | 0.538*** | 0.189* |
| | (0.057) | (0.097) |
| | [0.000] | [0.108] |
| R-squared | 0.353 | 0.183 |
| Number of observations | 2,785 | 2,304 |
| Mean | 0.555 | 0.380 |
| Control group mean | 0.272 | 0.230 |
| **P-values on test of equality:** | | |
| Partial subsidy = Full subsidy | 0.097 | 0.525 |
| 1/3 subsidy = 2/3 subsidy | 0.196 | 0.604 |
| 1/3 subsidy = Full subsidy | 0.016 | 0.807 |
| 2/3 subsidy = Full subsidy | 0.339 | 0.481 |

Notes: All regressions include a standard set of covariates (individual, household, and community) and baseline measure of dependent variable. *P*-values for the equality of effect estimates for various pairs of treatment groups are also presented. Robust standard errors clustered at community level are reported in parentheses. Wild-cluster bootstrap-t *p*-values are reported in square brackets. *, **, and *** denote statistical significance at 10 %, 5 %, and 1 % levels, respectively.

## Table 3: Characteristics of Compliers, Always Takers, and Never Takers

| | Short-run | | | | | | | Long-run | | | | | | |
|---|---|---|---|---|---|---|---|---|---|---|---|---|---|---|
| | Mean | | | | t-stat | | | Mean | | | | t-stat | | |
| | Total | Complier | Always | Never | C=A | C=N | A=N | Total | Complier | Always | Never | C=A | C=N | A=N |
| | (1) | (2) | (3) | (4) | (5) | (6) | (7) | (8) | (9) | (10) | (11) | (12) | (13) | (14) |
| ***Proportion*** | 100 | 47.4 | 27.1 | 25.5 | | | | 100 | 24.3 | 23.0 | 52.7 | | | |
| ***Panel A: Individual Characteristics*** | | | | | | | | | | | | | | |
| Age | 23.78 | 24.34 | 20.48 | 24.39 | 3.61 | -0.05 | -2.58 | 23.78 | 18.90 | 21.46 | 27.08 | -1.79 | -10.12 | -3.43 |
| Male | 0.48 | 0.51 | 0.47 | 0.48 | 1.18 | 1.17 | -0.15 | 0.48 | 0.51 | 0.44 | 0.51 | 1.90 | -0.04 | -1.71 |
| Christian | 0.42 | 0.43 | 0.51 | 0.40 | -2.74 | 1.24 | 2.90 | 0.42 | 0.37 | 0.53 | 0.42 | -4.58 | -2.38 | 3.00 |
| Dagaaba (ethnic group) | 0.52 | 0.61 | 0.54 | 0.44 | 2.14 | 6.89 | 2.79 | 0.52 | 0.59 | 0.53 | 0.51 | 1.70 | 4.39 | 0.51 |
| Has some formal education | 0.34 | 0.38 | 0.35 | 0.26 | 0.81 | 5.61 | 2.78 | 0.34 | 0.42 | 0.35 | 0.30 | 2.16 | 7.61 | 1.43 |
| Has a health condition (≥ 6 months) | 0.07 | 0.08 | 0.05 | 0.06 | 2.09 | 1.37 | -0.61 | 0.07 | 0.04 | 0.07 | 0.09 | -1.57 | -4.62 | -0.96 |
| Probably sick next year | 0.44 | 0.44 | 0.47 | 0.45 | -1.99 | -1.56 | 0.62 | 0.44 | 0.41 | 0.46 | 0.45 | -2.62 | -4.34 | 0.56 |
| Illness | | | | | | | | | | | | | | |
| Ill in the last four weeks | 0.12 | 0.13 | 0.11 | 0.14 | 1.21 | -0.13 | -0.98 | 0.12 | 0.14 | 0.17 | 0.11 | -1.17 | 3.05 | 2.26 |
| No. of days ill in the last four weeks | 0.92 | 1.03 | 0.75 | 0.89 | 1.62 | 0.91 | -0.58 | 0.92 | 0.85 | 1.05 | 0.93 | -0.73 | -0.57 | 0.38 |
| Could not do normal activities in the last four weeks | 0.08 | 0.10 | 0.04 | 0.08 | 5.80 | 1.93 | -2.28 | 0.08 | 0.10 | 0.10 | 0.07 | 0.23 | 3.01 | 1.04 |
| No. of days could not perform normal activities in the last four week | 0.54 | 0.74 | 0.33 | 0.43 | 3.39 | 3.96 | -0.69 | 0.54 | 0.42 | 0.68 | 0.57 | -1.18 | -1.39 | 0.47 |
| Illness due to Malaria | | | | | | | | | | | | | | |
| Ill in the last four weeks | 0.05 | 0.05 | 0.05 | 0.04 | -0.40 | 0.63 | 0.70 | 0.05 | 0.07 | 0.08 | 0.03 | -0.48 | 7.78 | 2.69 |
| No. of days ill in the last four weeks | 0.24 | 0.29 | 0.32 | 0.16 | -0.30 | 3.12 | 1.63 | 0.24 | 0.34 | 0.53 | 0.14 | -1.05 | 4.16 | 2.05 |
| Could not do normal activities in the last four weeks | 0.03 | 0.04 | 0.02 | 0.02 | 3.23 | 2.16 | -0.71 | 0.03 | 0.07 | 0.03 | 0.01 | 3.64 | 14.82 | 1.29 |
| No. of days could not perform normal activities in the last four week | 0.15 | 0.17 | 0.16 | 0.13 | 0.16 | 1.05 | 0.40 | 0.15 | 0.30 | 0.31 | 0.05 | -0.10 | 13.68 | 1.51 |
| Visited health facility in the last four weeks | 0.04 | 0.04 | 0.05 | 0.03 | -0.43 | 1.52 | 1.14 | 0.04 | 0.04 | 0.06 | 0.03 | -1.31 | 1.53 | 1.74 |
| Visited health facility in the last six months | 0.07 | 0.07 | 0.09 | 0.06 | -1.16 | 0.73 | 1.37 | 0.07 | 0.05 | 0.13 | 0.05 | -3.40 | -0.26 | 3.12 |
| Number of visits in the last four weeks | 0.07 | 0.06 | 0.08 | 0.05 | -0.80 | 0.81 | 1.07 | 0.07 | 0.02 | 0.10 | 0.06 | -2.24 | -2.78 | 0.99 |
| Visited health facility in the last four weeks for malaria treatment | 0.01 | 0.01 | 0.01 | 0.01 | 0.43 | 1.73 | 0.50 | 0.01 | 0.01 | 0.03 | 0.00 | -1.19 | 8.31 | 2.13 |
| Made out of pocket expense in the last six months | 0.14 | 0.15 | 0.12 | 0.14 | 1.55 | 0.89 | -0.57 | 0.14 | 0.15 | 0.16 | 0.13 | -0.20 | 1.99 | 1.03 |
| Ever enrolled in NHIS | 0.36 | 0.30 | 0.65 | 0.30 | -12.98 | 0.23 | 10.02 | 0.36 | 0.40 | 0.41 | 0.40 | -0.22 | -0.25 | 0.08 |
| Currently enrolled in NHIS | 0.20 | 0.05 | 0.47 | 0.15 | -14.99 | -6.38 | 9.39 | 0.20 | 0.14 | 0.30 | 0.17 | -5.09 | -2.34 | 3.79 |
| Slept under mosquito nets (12 years old or older) | 0.50 | 0.60 | 0.45 | 0.49 | 3.64 | 3.27 | -0.80 | 0.50 | 0.62 | 0.53 | 0.51 | 1.76 | 4.80 | 0.44 |
| ***Panel B: Household Characteristics*** | | | | | | | | | | | | | | |
| HH Size | 8.70 | 9.35 | 8.34 | 8.57 | 5.11 | 5.41 | -0.92 | 8.70 | 10.18 | 7.99 | 8.86 | 12.989 | 12.29 | -4.07 |
| Number of children under 18 | 5.14 | 5.42 | 5.17 | 5.08 | 1.37 | 2.77 | 0.39 | 5.14 | 5.84 | 5.12 | 5.26 | 3.2613 | 6.10 | -0.57 |
| Male head HH | 0.83 | 0.84 | 0.87 | 0.87 | -0.46 | -0.54 | 0.02 | 0.83 | 0.76 | 0.92 | 0.85 | -2.699 | -2.98 | 1.04 |
| Owns farming land | 0.55 | 0.69 | 0.48 | 0.53 | 7.08 | 6.45 | -1.24 | 0.55 | 0.79 | 0.48 | 0.56 | 8.6766 | 13.45 | -1.94 |
| Owns mosquito net | 0.71 | 0.74 | 0.73 | 0.67 | 0.39 | 2.86 | 1.37 | 0.71 | 0.52 | 0.91 | 0.74 | -18.36 | -18.15 | 6.04 |
| Knowledge about NHIS | 0.59 | 0.57 | 0.61 | 0.61 | -2.00 | -2.34 | 0.00 | 0.59 | 0.59 | 0.60 | 0.59 | -0.351 | -0.21 | 0.25 |
| Household assets (principal component score) | 0.60 | 1.09 | 0.77 | 0.59 | 2.56 | 5.69 | 1.19 | 0.60 | 2.29 | 0.16 | 0.59 | 23.213 | 71.18 | -3.50 |

Note: This table presents the mean individual (Panel A) and household (Panel B) characteristics of the entire sample, compliers and always takers, and never takers. The mean characteristics of compliers are estimated from Equation (4). Columns 5-7 and 12-14 present the t-statistics from the two-sample t-test comparing compliers with always takers, compliers with never takers, and always takers with never takers, respectively.

## Table 4: Selection by Subsidy Level

| Sample | Among those enrolled in the short run | | | | |
|---|---|---|---|---|---|
| Independent variable: Received full subsidy | Coefficient | Standard error | bootstrap *p*-values | N | R-squared |
| | (1) | (2) | (3) | (4) | (5) |
| **Panel A: Short run** | | | | | |
| Healthy or very healthy | -0.022 | (0.027) | 0.376 | 413 | 0.037 |
| # Days ill last month | -0.034 | (0.092) | 0.775 | 1,238 | 0.005 |
| Could not perform normal daily activities due to illness last month | -0.007 | (0.019) | 0.741 | 1,244 | 0.010 |
| # days could not perform normal daily activities in the last month | -0.008 | (0.262) | 0.978 | 1,244 | 0.003 |
| # Days ill last month (Malaria) | -0.045 | (0.034) | 0.257 | 1,237 | 0.011 |
| Could not perform normal daily activities due to illness last month (Malaria) | -0.006 | (0.011) | 0.587 | 1,238 | 0.009 |
| # days could not perform normal daily activities in the last month (Malaria) | -0.031 | (0.107) | 0.800 | 1,238 | 0.003 |
| Visited health facility in last four weeks | -0.010 | (0.017) | 0.600 | 1,152 | 0.017 |
| Visited health facility in last six months | -0.001 | (0.027) | 0.978 | 1,223 | 0.025 |
| # of visits in last six months | 0.004 | (0.012) | 0.794 | 1,148 | 0.010 |
| Visited Facility for malaria treatment in the last four weeks | -0.018 | (0.011) | 0.169 | 1,200 | 0.008 |
| Made an out-of-pocket for health service in the last six months | -0.006 | (0.016) | 0.754 | 1,244 | 0.008 |
| Standardized treatment effects (health status) | -0.003 | (0.005) | | 7,852 | 0.006 |
| Standardized treatment effects (health care utilization) | -0.007 | (0.008) | | 5,967 | 0.009 |
| Sample | Among those enrolled in the long run | | | | |
| Independent variable: Received full subsidy | Coefficient | Standard error | bootstrap *p*-values | N | R-squared |
| | (1) | (2) | (3) | (4) | (5) |
| **Panel B: Long run** | | | | | |
| Healthy or very healthy | 0.210* | (0.103) | 0.117 | 174 | 0.078 |
| # Days ill last month | -1.106*** | (0.338) | 0.012 | 674 | 0.049 |
| Could not perform normal daily activities due to illness last month | -0.085* | (0.044) | 0.117 | 674 | 0.027 |
| # days could not perform normal daily activities in the last month | -0.730* | (0.411) | 0.066 | 674 | 0.033 |
| # Days ill last month (Malaria) | -0.613*** | (0.206) | 0.032 | 674 | 0.037 |
| Could not perform normal daily activities due to illness last month (Malaria) | -0.079** | (0.033) | 0.034 | 674 | 0.034 |
| # days could not perform normal daily activities in the last month (Malaria) | -0.522** | (0.222) | 0.023 | 674 | 0.037 |
| Visited health facility in last four weeks | -0.122*** | (0.034) | 0.018 | 674 | 0.044 |
| Visited health facility in last six months | -0.242*** | (0.075) | 0.033 | 674 | 0.088 |
| # of visits in last six months | -0.099** | (0.041) | 0.080 | 674 | 0.033 |
| Visited Facility for malaria treatment in the last four weeks | -0.090** | (0.037) | 0.073 | 674 | 0.033 |
| Made an out-of-pocket for health service in the last six months | -0.034 | (0.023) | 0.152 | 674 | 0.018 |
| Standardized treatment effects (health status) | -0.076** | (0.031) | | 4,218 | 0.032 |
| Standardized treatment effects (health care utilization) | -0.116*** | (0.034) | | 3,370 | 0.038 |

Notes: This table reports estimation results of running regression of each selected health characteristics on an indicator variable that takes value of one if receiving full subsidy (zero price) and zero if receiving partial subsidy (positive price). We control for indicators of other interventions involving subsidy: *Subsidy + Campaign, Subsidy + Convenience,* and *Subsidy + Campaign + Convenience.* Sample is restricted to those who received partial and full subsidy. Panel A summarizes regression results when sample is restricted to those who enrolled in the short run. Panel B summarizes results when sample is restricted to those who enrolled in the long run. Standardized treatment effects on health status and health care utilization in the short and long run are reported in the last two rows of Panels A and B, respectively. Robust standard errors clustered at community level reported in parantheses. Wild-cluster bootstrap-t *p*-values are reported in Column 3. *, **, and *** denote statistical significance at 10 %, 5 %, and 1 % level respectively.

## Table 5: Effects on Healthcare Services Utilization (Short Run)

| | Short run | | | | | |
|---|---|---|---|---|---|---|
| | Visited health facility in last four weeks | Visited health facility in last six months | # of visits in last four weekss | Visited facility for malaria treatment in the last four weeks | Made out-of-pocket for health service in the last six months | Standardized treatment effects |
| | (1) | (2) | (3) | (4) | (5) | (6) |
| **Panel A: ITT results** | | | | | | |
| **Panel A1** | | | | | | |
| Any subsidy | 0.020 | 0.053** | 0.031 | 0.018** | 0.011 | 0.021* |
| | (0.017) | (0.021) | (0.022) | (0.008) | (0.013) | (0.011) |
| | [0.294] | [0.035] | [0.247] | [0.069] | [0.457] | |
| | {0.500} | {0.202} | {0.475} | {0.254} | {0.500} | |
| R-squared | 0.099 | 0.119 | 0.062 | 0.066 | 0.092 | 0.054 |
| **Panel A2** | | | | | | |
| Partial subsidy | -0.008 | 0.009 | -0.003 | 0.008 | -0.009 | -0.0001 |
| | (0.013) | (0.019) | (0.024) | (0.009) | (0.015) | (0.010) |
| | [0.591] | [0.623] | [0.883] | [0.387] | [0.598] | |
| | {0.944} | {0.944} | {0.944} | {0.869} | {0.944} | |
| Full subsidy | 0.008 | 0.001 | 0.007 | 0.001 | -0.013 | 0.004 |
| | (0.022) | (0.031) | (0.038) | (0.012) | (0.023) | (0.016) |
| | [0.716] | [0.963] | [0.875] | [0.973] | [0.585] | |
| | {0.990} | {0.998} | {0.998} | {0.998} | {0.979} | |
| R-squared | 0.106 | 0.129 | 0.065 | 0.074 | 0.094 | 0.058 |
| **Panel A3** | | | | | | |
| 1/3 subsidy | -0.0003 | 0.009 | -0.012 | 0.009 | -0.015 | 0.001 |
| | (0.016) | (0.022) | (0.023) | (0.011) | (0.014) | (0.011) |
| | [0.974] | [0.668] | [0.629] | [0.464] | [0.325] | |
| | {0.986} | {0.918} | {0.918} | {0.881} | {0.803} | |
| 2/3 subsidy | -0.014 | 0.009 | 0.004 | 0.007 | -0.004 | -0.001 |
| | (0.014) | (0.023) | (0.030) | (0.010) | (0.020) | (0.011) |
| | [0.390] | [0.697] | [0.901] | [0.480] | [0.842] | |
| | {0.851} | {0.965} | {0.974} | {0.898} | {0.974} | |
| Full subsidy | 0.007 | 0.001 | 0.008 | 0.0004 | -0.012 | 0.004 |
| | (0.021) | (0.031) | (0.038) | (0.011) | (0.023) | (0.016) |
| | [0.746] | [0.962] | [0.830] | [0.970] | [0.678] | |
| | {0.996} | {0.998} | {0.996} | {0.998} | {0.984} | |
| R-squared | 0.106 | 0.129 | 0.065 | 0.074 | 0.094 | 0.058 |
| Number of observations | 2,130 | 2,710 | 2,124 | 2,252 | 2,805 | 11,008 |
| **Panel B: 2SLS results** | | | | | | |
| Enrolled in NHIS | -0.007 | 0.011 | 0.005 | 0.011 | -0.019 | 0.002 |
| | (0.024) | (0.041) | (0.050) | (0.015) | (0.035) | (0.021) |
| First-stage F-statistics | 26.646 | 25.134 | 25.559 | 26.421 | 26.608 | 27.884 |
| Control group mean | 0.038 | 0.102 | 0.032 | 0.019 | 0.046 | -0.011 |
| **P-values on test of equality:** | | | | | | |
| Partial subsidy = Full subsidy | 0.491 | 0.751 | 0.783 | 0.543 | 0.825 | 0.784 |
| 1/3 subsidy = 2/3 subsidy | 0.419 | 0.987 | 0.500 | 0.893 | 0.515 | 0.819 |
| 1/3 subsidy = Full subsidy | 0.743 | 0.786 | 0.560 | 0.480 | 0.821 | 0.854 |
| 2/3 subsidy = Full subsidy | 0.437 | 0.769 | 0.908 | 0.625 | 0.710 | 0.770 |

Notes: Panels A and B report ITT and 2SLS results, respectively. Panels A1, A2, and A3 report the effects of receiving any subsidy, partial and full subsidy, and each subsidy level (1/3, 2/3, and full), respectively. All regressions include a standard set of covariates (individual, household, and community) and baseline measure of dependent variable. Standardized treatment effects are reported in Column 6. *P*-values for the equality of effect estimates for various pairs of treatment groups are also presented. Robust standard errors clustered at community level are reported in parentheses. Wild-cluster bootstrap-t *p*-values are reported in square brackets. Family-wise *p*-values are reported in curly brackets. *, **, and *** denote statistical significance at 10 %, 5 %, and 1 % levels, respectively.

## Table 6: Effects on Healthcare Services Utilization (Long Run)

| | Long run | | | | | |
|---|---|---|---|---|---|---|
| | Visited health facility in last four weeks | Visited health facility in last six months | # of visits in last four weekss | Visited facility for malaria treatment in the last four weeks | Made out-of-pocket for health service in the last six months | Standardized treatment effects |
| | (1) | (2) | (3) | (4) | (5) | (6) |
| **Panel A: ITT results** | | | | | | |
| **Panel A1** | | | | | | |
| Any subsidy | 0.038*** | 0.079*** | 0.033*** | 0.028*** | 0.009 | 0.038*** |
| | (0.009) | (0.018) | (0.010) | (0.010) | (0.008) | (0.010) |
| | [0.000] | [0.000] | [0.009] | [0.022] | [0.378] | |
| | {0.018} | {0.012} | {0.054} | {0.077} | {0.375} | |
| R-squared | 0.077 | 0.084 | 0.060 | 0.064 | 0.087 | 0.062 |
| **Panel A2** | | | | | | |
| Partial subsidy | 0.048*** | 0.108*** | 0.038*** | 0.035** | 0.013 | 0.048*** |
| | (0.013) | (0.025) | (0.012) | (0.014) | (0.010) | (0.012) |
| | [0.001] | [0.001] | [0.004] | [0.024] | [0.256] | |
| | {0.062} | {0.033} | {0.08} | {0.151} | {0.287} | |
| Full subsidy | -0.029 | -0.013 | -0.033 | -0.035 | -0.038* | -0.037 |
| | (0.020) | (0.053) | (0.023) | (0.021) | (0.022) | (0.023) |
| | [0.222] | [0.849] | [0.243] | [0.201] | [0.057] | |
| | {0.63} | {0.854} | {0.63} | {0.63} | {0.63} | |
| R-squared | 0.094 | 0.102 | 0.078 | 0.084 | 0.103 | 0.079 |
| **Panel A3** | | | | | | |
| 1/3 subsidy | 0.020 | 0.085** | 0.016 | 0.019 | 0.025 | 0.036* |
| | (0.015) | (0.034) | (0.017) | (0.016) | (0.027) | (0.021) |
| | [0.297] | [0.057] | [0.441] | [0.328] | [0.719] | |
| | {0.603} | {0.41} | {0.681} | {0.615} | {0.681} | |
| 2/3 subsidy | 0.070*** | 0.125*** | 0.056*** | 0.048*** | 0.004 | 0.058*** |
| | (0.016) | (0.032) | (0.017) | (0.018) | (0.011) | (0.015) |
| | [0.000] | [0.001] | [0.010] | [0.019] | [0.794] | |
| | {0.029} | {0.041} | {0.058} | {0.114} | {0.812} | |
| Full subsidy | -0.025 | -0.010 | -0.030 | -0.033 | -0.039 | -0.035 |
| | (0.020) | (0.054) | (0.024) | (0.021) | (0.024) | (0.024) |
| | [0.319] | [0.913] | [0.265] | [0.249] | [0.100] | |
| | {0.647} | {0.894} | {0.647} | {0.647} | {0.647} | |
| R-squared | 0.099 | 0.103 | 0.081 | 0.086 | 0.105 | 0.080 |
| Number of observations | 2,228 | 2,688 | 2,231 | 2,228 | 2,688 | 11,140 |
| **Panel B: 2SLS results** | | | | | | |
| Enrolled in NHIS | 0.051 | 0.145** | 0.038 | 0.031 | -0.017 | 0.041 |
| | (0.033) | (0.061) | (0.031) | (0.033) | (0.020) | (0.032) |
| First-stage F-statistics | 33.381 | 34.796 | 32.094 | 31.355 | 32.844 | 35.857 |
| Control group mean | 0.017 | 0.050 | 0.036 | 0.010 | 0.013 | -0.021 |
| **P-values on test of equality:** | | | | | | |
| Partial subsidy = Full subsidy | 0.000 | 0.010 | 0.004 | 0.002 | 0.085 | 0.001 |
| 1/3 subsidy = 2/3 subsidy | 0.015 | 0.355 | 0.110 | 0.177 | 0.539 | 0.417 |
| 1/3 subsidy = Full subsidy | 0.068 | 0.092 | 0.093 | 0.049 | 0.197 | 0.063 |
| 2/3 subsidy = Full subsidy | 0.000 | 0.006 | 0.003 | 0.001 | 0.019 | 0.000 |

Notes: Panels A and B report ITT and 2SLS results, respectively. Panels A1, A2, and A3 report the effects of receiving any subsidy, partial and full subsidy, and each subsidy level (1/3, 2/3, and full), respectively. All regressions include a standard set of covariates (individual, household, and community) and baseline measure of dependent variable. Standardized treatment effects are reported in Column 6. *P*-values for the equality of effect estimates for various pairs of treatment groups are also presented. Robust standard errors clustered at community level are reported in parentheses. Wild-cluster bootstrap-t *p*-values are reported in square brackets. Family-wise *p*-values are reported in curly brackets. *, **, and *** denote statistical significance at 10 %, 5 %, and 1 % levels, respectively.

## Table 7: Effects on Health Status (Short Run)

| | Short run | | | | |
|---|---|---|---|---|---|
| | Healthy or very healthy | # of days ill last four weeks | Could not perform normal daily activities due to illness last four weeks | # of days could not perform normal daily activities due to illness in the last four weeks | Standardized treatment effects |
| | (1) | (2) | (3) | (4) | (5) |
| **Panel A: ITT results** | | | | | |
| **Panel A1** | | | | | |
| Any subsidy | 0.059 | -0.275** | 0.007 | -0.355 | -0.018 |
| | (0.042) | (0.116) | (0.017) | (0.280) | (0.013) |
| | [0.182] | [0.019] | [0.743] | [0.289] | |
| | {0.501} | {0.179} | {0.741} | {0.501} | |
| R-squared | 0.176 | 0.085 | 0.079 | 0.092 | 0.063 |
| **Panel A2** | | | | | |
| Partial subsidy | 0.130*** | -0.309** | -0.012 | -0.076 | -0.025* |
| | (0.037) | (0.135) | (0.017) | (0.368) | (0.013) |
| | [0.003] | [0.036] | [0.567] | [0.850] | |
| | {0.035} | {0.145} | {0.717} | {0.865} | |
| Full subsidy | 0.118** | -0.409* | -0.021 | -0.511 | -0.041* |
| | (0.044) | (0.211) | (0.030) | (0.508) | (0.022) |
| | [0.009] | [0.071] | [0.527] | [0.357] | |
| | {0.099} | {0.222} | {0.543} | {0.488} | |
| R-squared | 0.192 | 0.086 | 0.080 | 0.094 | 0.064 |
| **Panel A3** | | | | | |
| 1/3 subsidy | 0.121*** | -0.399** | -0.011 | -0.383 | -0.034** |
| | (0.041) | (0.158) | (0.023) | (0.411) | (0.017) |
| | [0.011] | [0.033] | [0.68] | [0.408] | |
| | {0.071} | {0.094} | {0.688} | {0.546} | |
| 2/3 subsidy | 0.136*** | -0.241 | -0.013 | 0.166 | -0.018 |
| | (0.044) | (0.174) | (0.020) | (0.429) | (0.015) |
| | [0.011] | [0.245] | [0.623] | [0.773] | |
| | {0.09} | {0.54} | {0.741} | {0.757} | |
| Full subsidy | 0.119*** | -0.395* | -0.021 | -0.453 | -0.040* |
| | (0.044) | (0.214) | (0.030) | (0.508) | (0.022) |
| | [0.01] | [0.115] | [0.534] | [0.425] | |
| | {0.097} | {0.256} | {0.554} | {0.554} | |
| R-squared | 0.192 | 0.086 | 0.080 | 0.095 | 0.064 |
| Number of observations | 861 | 2,768 | 2,775 | 2,677 | 8,824 |
| **Panel B: 2SLS results** | | | | | |
| Enrolled in NHIS | 0.238*** | -0.691** | -0.039 | -0.536 | -0.067** |
| | (0.074) | (0.311) | (0.037) | (0.806) | (0.028) |
| First-stage F-statistics | 22.737 | 26.609 | 27.019 | 26.858 | 27.464 |
| Control group mean | 0.817 | 0.617 | 0.081 | 1.379 | -0.019 |
| **P-values on test of equality:** | | | | | |
| Partial subsidy = Full subsidy | 0.723 | 0.477 | 0.710 | 0.179 | 0.278 |
| 1/3 subsidy = 2/3 subsidy | 0.722 | 0.409 | 0.921 | 0.212 | 0.400 |
| 1/3 subsidy = Full subsidy | 0.970 | 0.979 | 0.701 | 0.848 | 0.731 |
| 2/3 subsidy = Full subsidy | 0.664 | 0.333 | 0.764 | 0.100 | 0.221 |

Notes: Panels A and B report ITT and 2SLS results, respectively. Panels A1, A2, and A3 report the effects of receiving any subsidy, partial and full subsidy, and each subsidy level (1/3, 2/3, and full), respectively. All regressions include a standard set of covariates (individual, household, and community) and baseline measure of dependent variable. Standardized treatment effects are reported in Column 5. *P*-values for the equality of effect estimates for various pairs of treatment groups are also presented. Robust standard errors clustered at community level are reported in parentheses. Wild-cluster bootstrap-t *p*-values are reported in square brackets. Family-wise *p*-values are reported in curly brackets. *, **, and *** denote statistical significance at 10 %, 5 %, and 1 % levels, respectively.

Table 8: Effects on Health Status (Long Run)

| | Long run | | | | |
|---|---|---|---|---|---|
| | Healthy or very healthy | # of days ill last four weeks | Could not perform normal daily activities due to illness last four weeks | # of days could not perform normal daily activities due to illness in the last four weeks | Standardized treatment effects |
| | (1) | (2) | (3) | (4) | (5) |
| **Panel A: ITT results** | | | | | |
| **Panel A1** | | | | | |
| Any subsidy | -0.076 | 0.221** | 0.030*** | 0.178** | 0.036*** |
| | (0.050) | (0.092) | (0.010) | (0.074) | (0.012) |
| | [0.166] | [0.035] | [0.009] | [0.028] | |
| | {0.194} | {0.132} | {0.055} | {0.132} | |
| R-squared | 0.289 | 0.071 | 0.084 | 0.059 | 0.054 |
| **Panel A2** | | | | | |
| Partial subsidy | -0.156** | 0.294*** | 0.048*** | 0.284*** | 0.058*** |
| | (0.058) | (0.087) | (0.013) | (0.082) | (0.013) |
| | [0.166] | [0.035] | [0.009] | [0.028] | |
| | {0.051} | {0.031} | {0.029} | {0.031} | |
| Full subsidy | -0.130 | -0.352* | -0.006 | -0.179 | -0.028 |
| | (0.093) | (0.175) | (0.020) | (0.185) | (0.027) |
| | [0.042] | [0.025] | [0.007] | [0.024] | |
| | {0.566} | {0.379} | {0.82} | {0.604} | |
| R-squared | 0.301 | 0.083 | 0.099 | 0.077 | 0.068 |
| **Panel A3** | | | | | |
| 1/3 subsidy | -0.081 | 0.205 | 0.035** | 0.240* | 0.042** |
| | (0.067) | (0.141) | (0.016) | (0.129) | (0.020) |
| | [0.254] | [0.127] | [0.823] | [0.535] | |
| | {0.396} | {0.396} | {0.286} | {0.325} | |
| 2/3 subsidy | -0.221*** | 0.362*** | 0.058*** | 0.318*** | 0.071*** |
| | (0.074) | (0.100) | (0.016) | (0.108) | (0.018) |
| | [0.29] | [0.212] | [0.063] | [0.123] | |
| | {0.051} | {0.032} | {0.033} | {0.051} | |
| Full subsidy | -0.140 | -0.339* | -0.004 | -0.171 | -0.026 |
| | (0.090) | (0.170) | (0.020) | (0.176) | (0.026) |
| | [0.021] | [0.008] | [0.017] | [0.033] | |
| | {0.497} | {0.387} | {0.872} | {0.614} | |
| R-squared | 0.307 | 0.084 | 0.100 | 0.077 | 0.068 |
| Number of observations | 658 | 2,666 | 2,661 | 2,564 | 8,309 |
| **Panel B: 2SLS results** | | | | | |
| Enrolled in NHIS | -0.306** | 0.173 | 0.067** | 0.289 | 0.066* |
| | (0.120) | (0.226) | (0.031) | (0.202) | (0.036) |
| First-stage F-statistics | 42.373 | 34.195 | 33.371 | 33.838 | 33.695 |
| Control group mean | 0.791 | 0.413 | 0.013 | 0.096 | 0.011 |
| **P-values on test of equality:** | | | | | |
| Partial subsidy = Full subsidy | 0.764 | 0.000 | 0.010 | 0.015 | 0.002 |
| 1/3 subsidy = 2/3 subsidy | 0.140 | 0.351 | 0.278 | 0.650 | 0.275 |
| 1/3 subsidy = Full subsidy | 0.397 | 0.009 | 0.089 | 0.015 | 0.012 |
| 2/3 subsidy = Full subsidy | 0.476 | 0.000 | 0.008 | 0.028 | 0.002 |

Notes: Panels A and B report ITT and 2SLS results, respectively. Panels A1, A2, and A3 report the effects of receiving any subsidy, partial and full subsidy, and each subsidy level (1/3, 2/3, and full), respectively. All regressions include a standard set of covariates (individual, household, and community) and baseline measure of dependent variable. Standardized treatment effects are reported in Column 5. *P*-values for the equality of effect estimates for various pairs of treatment groups are also presented. Robust standard errors clustered at community level are reported in parentheses. Wild-cluster bootstrap-t *p*-values are reported in square brackets. Family-wise *p*-values are reported in curly brackets. *, **, and *** denote statistical significance at 10 %, 5 %, and 1 % levels, respectively.

## Appendix A

Figure A.1. Wa West District Map

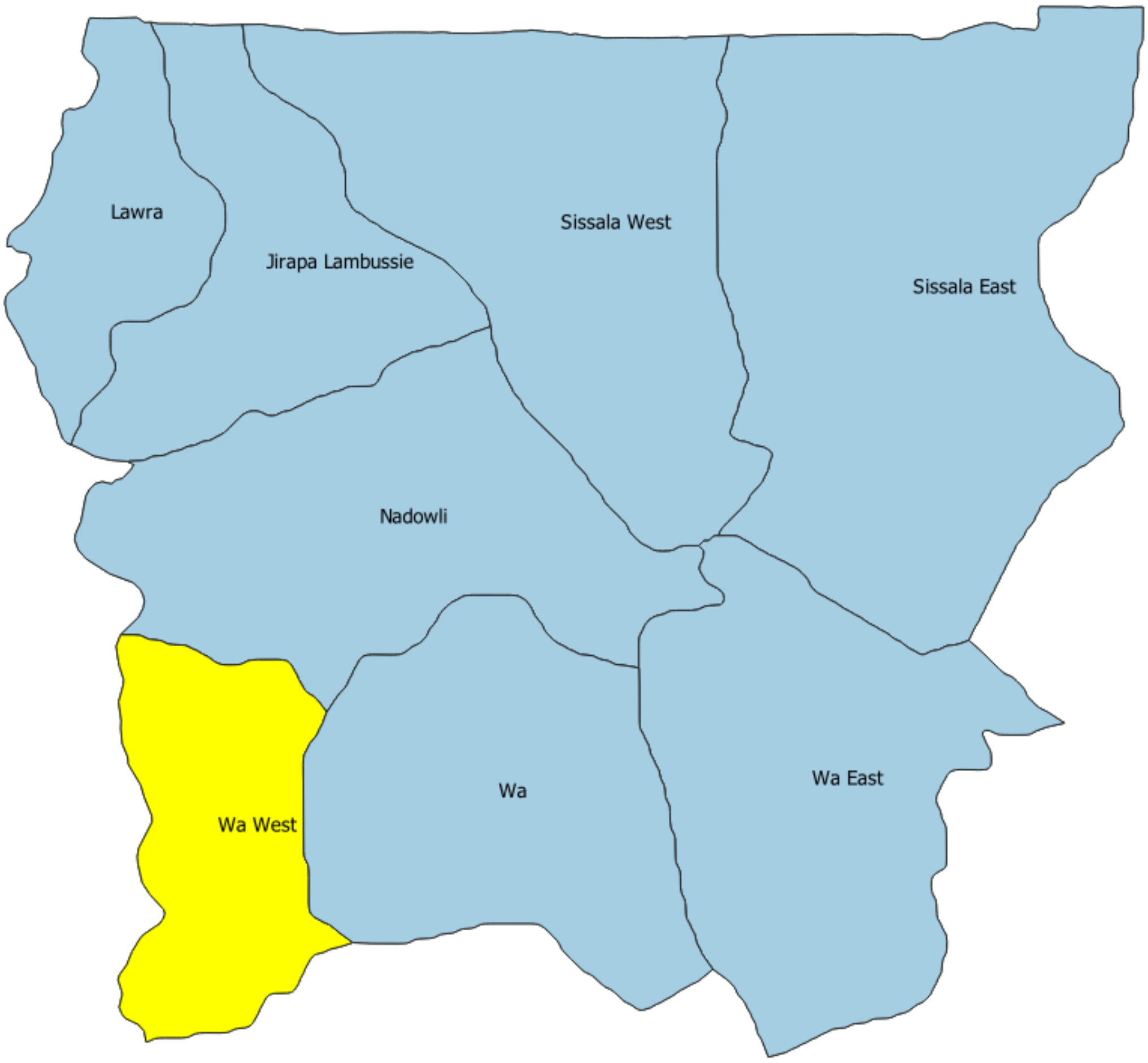

Note: This map shows the Upper West region of Ghana, which includes Wa West district (highlighted).

Table A1: NHIS Coverage

| Included Services | Exclusion List |
|---|---|
| 1 Out-Patient Services<br>i) General and specialized consultation and review<br>ii) Requested investigation (including laboratory investigations, x-rays and ultrasound scanning)<br>iii) Medication (prescription drugs on the NHIS Drug List)<br>iv) HIV/AIDS symptomatic treatment for opportunistic infection<br>v) Out-patient/Day Surgery Operations including hernia repairs, incision and drainage, hemorrhoidectomy<br>vi) Out-patient physiotherapy<br><br>2 In-Patient Services<br>i) General and specialist in-patient care<br>ii) Requested investigations<br>iii) Medication (prescription drugs on NHIS Drug List)<br>iv) Cervical and Breast Cancer Treatment<br>v) Surgical Operations<br>vi) In-patient physiotherapy<br>vii) Accommodation in general ward<br>viii) Feeding (where available)<br><br>3 Oral Health Services<br>i) Pain relief which includes incision and drainage, tooth extraction and temporary relief<br>ii) Dental restoration which includes simple amalgam fillings and temporary dressing<br><br>4 Eye Care Services<br>i) Refraction, visual fields and A-Scan<br>ii) Keratometry<br>iii) Cataract removal<br>iv) Eye lid surgery<br><br>5 Maternity Care<br>i) Antenatal care<br>ii) Deliveries (normal and assisted)<br>iii) Caesarian section<br>iv) Postnatal care<br><br>6 Emergencies<br>i) Medical emergencies<br>ii) Surgical emergencies including brain surgery due to accidents<br>iii) Pediatric emergencies<br>iv) Obstetric and gynecological emergencies<br>v) Road traffic accidents<br>vi) Industrial and workplace accidents<br>vii) Dialysis for acute renal failure | 1 Rehabilitation other than physiotherapy<br><br>2 Appliances and protheses including optical aids, hearing aids, othopedic aids and dentures<br><br>3 Cosmetic surgeries and aesthetic treatment<br><br>4 HIV retroviral drugs<br><br>5 Assisted reproduction eg artificial insemination and gynecological hormone replacement therapy<br><br>6 Echocardiography<br><br>7 Photography<br><br>8 Angiography<br><br>9 Orthotics<br><br>10 Dialysis for chronic renal failure<br><br>11 Heart and brain surgery other than those resulting from accident<br><br>12 Cancer treatment other than cervical ad breast cancer<br><br>13 Organ transplating<br><br>14 All drugs that not listed on the NHIS Drug List<br><br>15 Diagnosis and treatment abroad<br><br>16 Medical examinations for purposes of visa applications, Campaign and institutional driving license<br><br>17 VIP ward accommodation<br><br>18 Mortuary Services |

Source: NHIA (2011)

Table A2: Attrition

| | Short run | Long run |
|---|---|---|
| | (1) | (2) |
| **Panel A** | | |
| Any subsidy | 0.005 | -0.045 |
| | (0.021) | (0.036) |
| | [0.855] | [0.259] |
| R-squared | 0.133 | 0.102 |
| **Panel B** | | |
| Partial subsidy | 0.004 | -0.042 |
| | (0.025) | (0.038) |
| | [0.895] | [0.302] |
| Full subsidy | 0.013 | -0.065 |
| | (0.044) | (0.053) |
| | [0.809] | [0.311] |
| R-squared | 0.144 | 0.104 |
| **Panel C** | | |
| 1/3 subsidy | -0.005 | -0.039 |
| | (0.043) | (0.051) |
| | [0.924] | [0.533] |
| 2/3 subsidy | 0.011 | -0.045 |
| | (0.024) | (0.042) |
| | [0.664] | [0.314] |
| Full subsidy | 0.014 | -0.066 |
| | (0.044) | (0.051) |
| | [0.779] | [0.298] |
| R-squared | 0.144 | 0.104 |
| Mean | 0.05 | 0.21 |
| Number of observations | 2953 | 2953 |

Notes: Dependent variable is a binary variable indicating whether an individual had been attrited in the short- and long-run follow-up surveys. All regressions include a standard set of covariates (individual, household, and community). Robust standard errors clustered at community level reported in parantheses. *, **, and *** denote statistical significance at 10 %, 5 %, and 1 % levels, respectively.

## Table A3: Selective Retention of Health Insurance by Characteristics

| Sample | Among those enrolled in the baseline | | | | |
|---|---|---|---|---|---|
| Independent variable: Enrolled at the first follow-up | Coefficient | Standard error | bootstrap p-values | N | R-squared |
| | (1) | (2) | (3) | (4) | (5) |
| **Panel A: Short run** | | | | | |
| Healthy or very healthy | 0.020 | (0.058) | 0.740 | 161 | 0.001 |
| # Days ill last month | -0.354 | (0.338) | 0.338 | 531 | 0.004 |
| Could not perform normal daily activities due to illness last month | 0.029 | (0.038) | 0.507 | 535 | 0.002 |
| # days could not perform normal daily activities in the last month | -0.037 | (0.690) | 0.979 | 535 | 0.00001 |
| # Days ill last month (Malaria) | -0.064 | (0.119) | 0.617 | 531 | 0.001 |
| Could not perform normal daily activities due to illness last month (Malaria) | 0.017 | (0.019) | 0.444 | 532 | 0.002 |
| # days could not perform normal daily activities in the last month (Malaria) | -0.002 | (0.269) | 0.990 | 532 | 0.0000001 |
| Visited health facility in last four weeks | 0.035 | (0.025) | 0.219 | 497 | 0.004 |
| Visited health facility in last six months | 0.114** | (0.043) | 0.027 | 513 | 0.017 |
| # of visits in last six months | 0.037* | (0.021) | 0.121 | 494 | 0.004 |
| Visited Facility for malaria treatment in the last four weeks | 0.034* | (0.020) | 0.146 | 511 | 0.005 |
| Made an out-of-pocket for health service in the last six months | -0.010 | (0.029) | 0.849 | 535 | 0.001 |
| Standardized treatment effects (health status) | -0.001 | (0.017) | | 3,357 | 0.00001 |
| Standardized treatment effects (health care utilization) | 0.030* | (0.017) | | 2,550 | 0.004 |
| Sample | Among those enrolled in the short run | | | | |
| Independent variable: Enrolled at the second follow-up | Coefficient | Standard error | bootstrap p-values | N | R-squared |
| | (1) | (2) | (3) | (4) | (5) |
| **Panel B: Long run** | | | | | |
| Healthy or very healthy | -0.008 | (0.067) | 0.910 | 360 | 0.0001 |
| # Days ill last month | 0.210 | (0.168) | 0.253 | 1,305 | 0.003 |
| Could not perform normal daily activities due to illness last month | 0.021 | (0.017) | 0.239 | 1,305 | 0.003 |
| # days could not perform normal daily activities in the last month | 0.119 | (0.135) | 0.520 | 1,305 | 0.002 |
| # Days ill last month (Malaria) | 0.049 | (0.093) | 0.625 | 1,305 | 0.0003 |
| Could not perform normal daily activities due to illness last month (Malaria) | 0.016 | (0.015) | 0.305 | 1,305 | 0.002 |
| # days could not perform normal daily activities in the last month (Malaria) | 0.066 | (0.087) | 0.557 | 1,305 | 0.001 |
| Visited health facility in last four weeks | 0.047** | (0.018) | 0.003 | 1,305 | 0.012 |
| Visited health facility in last six months | 0.139*** | (0.039) | 0.000 | 1,305 | 0.042 |
| # of visits in last six months | 0.038** | (0.018) | 0.017 | 1,305 | 0.008 |
| Visited Facility for malaria treatment in the last four weeks | 0.030* | (0.016) | 0.044 | 1,305 | 0.006 |
| Made an out-of-pocket for health service in the last six months | -0.025* | (0.013) | 0.074 | 1,305 | 0.007 |
| Standardized treatment effects (health status) | 0.013 | (0.012) | | 8,190 | 0.001 |
| Standardized treatment effects (health care utilization) | 0.038** | (0.018) | | 6,525 | 0.006 |

Notes: This table reports estimation results of running univariate regression of each selected health characteristics on an enrollment indicator in short and long-run. Panel A summarizes regression results when sample is restricted to those who enrolled in the baseline. Panel B summarizes results when sample is restricted to those who enrolled in the short run. Standardized treatment effects on health status and health care utilization in the short and long run are reported in the last two rows of Panels A and B, respectively. Robust standard errors clustered at community level reported in parantheses. Robust standard errors clustered at community level reported in parantheses. Wild-cluster bootstrap-t *p*-values are reported in Column 3. *, **, and *** denote statistical significance at 10 %, 5 %, and 1 % level respectively.

Table A4: Effects on Health Behaviors.

| | Short run | Long run | | | |
|---|---|---|---|---|---|
| | Sleep under mosquito nets | Have mosquito nets | Sleep under mosquito nets | Water safe to drink | Standardized treatment effects |
| | (1) | (2) | (3) | (4) | (5) |
| **Panel A: ITT results** | | | | | |
| **Panel A1** | | | | | |
| Any subsidy | 0.085 | -0.035 | 0.016 | -0.072 | -0.032 |
| | (0.065) | (0.088) | (0.072) | (0.049) | (0.041) |
| | [0.251] | [0.752] | [0.85] | [0.218] | |
| | | {0.933} | {0.933} | {0.588} | |
| R-squared | 0.233 | 0.258 | 0.235 | 0.257 | 0.155 |
| **Panel A2** | | | | | |
| Partial subsidy | 0.098 | 0.039 | 0.036 | -0.071 | -0.018 |
| | (0.113) | (0.094) | (0.123) | (0.045) | (0.044) |
| | [0.457] | [0.758] | [0.806] | [0.149] | |
| | | {0.926} | {0.926} | {0.522} | |
| Full subsidy | 0.227* | -0.269** | -0.044 | -0.014 | -0.117** |
| | (0.118) | (0.106) | (0.118) | (0.068) | (0.051) |
| | [0.113] | [0.064] | [0.749] | [0.878] | |
| | | {0.241} | {0.946} | {0.946} | |
| R-squared | 0.247 | 0.318 | 0.259 | 0.275 | 0.179 |
| **Panel A3** | | | | | |
| 1/3 subsidy | 0.020 | 0.146 | 0.072 | -0.054 | 0.021 |
| | (0.110) | (0.117) | (0.127) | (0.057) | (0.047) |
| | [0.886] | [0.343] | [0.664] | [0.394] | |
| | | {0.716} | {0.724} | {0.724} | |
| 2/3 subsidy | 0.158 | -0.065 | 0.009 | -0.087* | -0.055 |
| | (0.141) | (0.089) | (0.131) | (0.044) | (0.049) |
| | [0.396] | [0.567] | [0.959] | [0.061] | |
| | | {0.796} | {0.956} | {0.355} | |
| Full subsidy | 0.238** | -0.294*** | -0.050 | -0.017 | -0.127** |
| | (0.118) | (0.097) | (0.120) | (0.068) | (0.051) |
| | [0.088] | [0.022] | [0.723] | [0.829] | |
| | | {0.143} | {0.931} | {0.931} | |
| R-squared | 0.252 | 0.333 | 0.260 | 0.276 | 0.182 |
| Number of observations | 1,422 | 1,101 | 1,092 | 497 | 2,069 |
| **Panel B: 2SLS results** | | | | | |
| Enrolled in NHIS | 0.306 | -0.184 | 0.027 | -0.091 | -0.085 |
| | (0.204) | (0.148) | (0.219) | (0.083) | (0.078) |
| First-stage F-statistics | 29.175 | 38.614 | 28.639 | 42.639 | 56.943 |
| Control group mean | 0.447 | 0.290 | 0.661 | 0.080 | 0.007 |
| **P-values on test of equality:** | | | | | |
| Partial subsidy = Full subsidy | 0.274 | 0.001 | 0.179 | 0.166 | 0.003 |
| 1/3 subsidy = 2/3 subsidy | 0.303 | 0.043 | 0.382 | 0.482 | 0.008 |
| 1/3 subsidy = Full subsidy | 0.096 | 0.0001 | 0.131 | 0.490 | 0.00004 |
| 2/3 subsidy = Full subsidy | 0.544 | 0.011 | 0.340 | 0.106 | 0.049 |

Note: Health behaviors are measured for those aged 12 years and above. Dependent variable in Column 4 is an indicator variable of whether a household member does anything to their water to make it safe to drink. Panels A and B report ITT and 2SLS results, respectively. Panels A1, A2, and A3 report the effects of receiving any subsidy, partial and full subsidy, and each subsidy level (1/3, 2/3, and full), respectively. All regressions include a standard set of covariates (individual, household, and community) and baseline measure of dependent variable. Standardized treatment effect in the long run is reported in Column 5. *P*-values for the equality of effect estimates for various pairs of treatment groups are also presented. Robust standard errors clustered at community level are reported in parentheses. Wild-cluster bootstrap-t *p*-values are reported in square brackets. Family-wise *p*-values are reported in curly brackets *, **, and *** denote statistical significance at 10 %, 5 %, and 1 % levels, respectively.

**Appendix B**

Figure B.1: Original Study Design

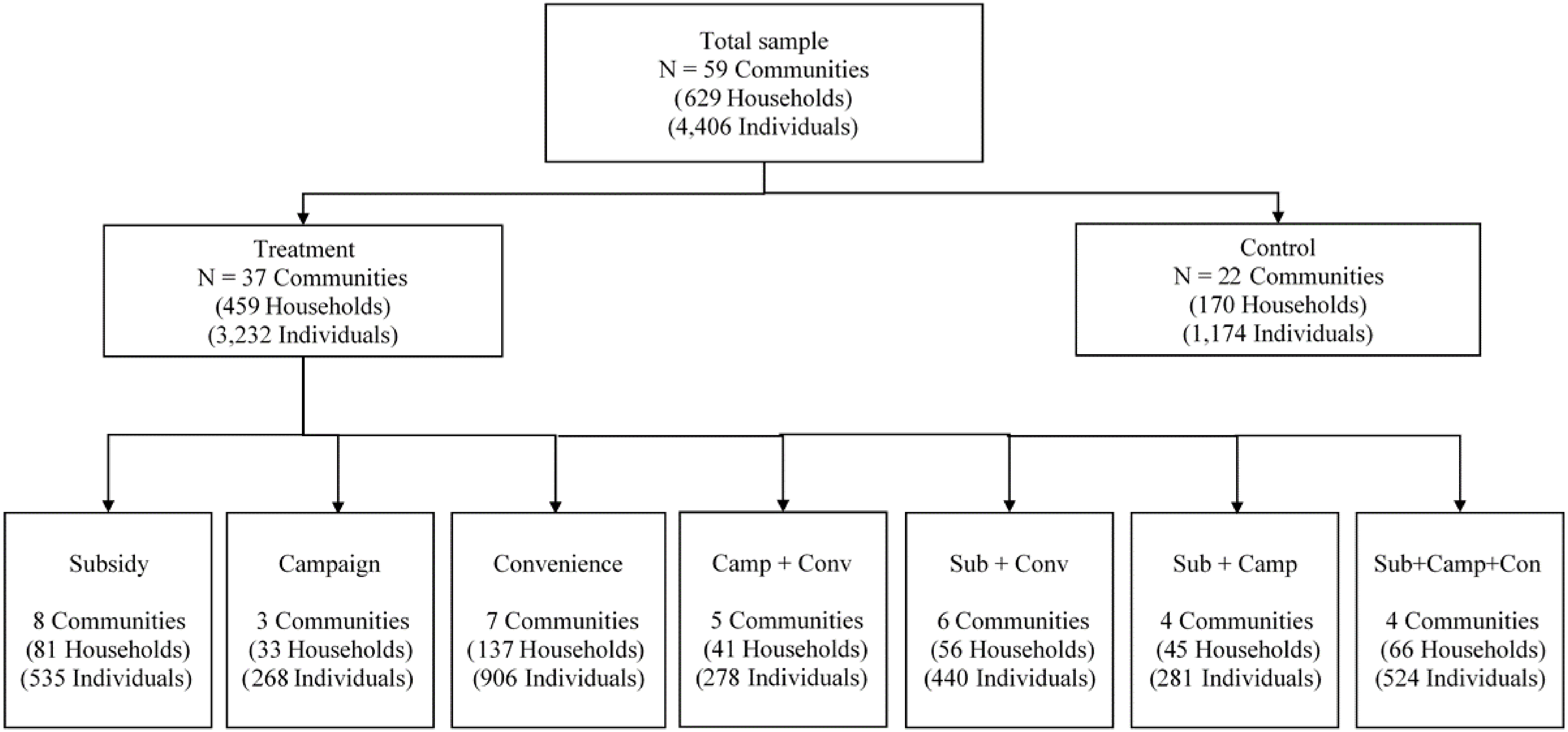

Table B.1: Effects of the Original Interventions on Enrollment.

| | Enrollment | |
|---|---|---|
| | Short-run | Long-run |
| | (1) | (2) |
| Subsidy only | 0.436*** | 0.160* |
| | (0.046) | (0.082) |
| Campaign only | 0.161** | 0.044 |
| | (0.080) | (0.066) |
| Convenience only | 0.007 | 0.195*** |
| | (0.066) | (0.072) |
| Campaign & Convenience | 0.231 | 0.182 |
| | (0.165) | (0.159) |
| Subsidy & Convenience | 0.347*** | 0.155** |
| | (0.078) | (0.064) |
| Subsidy & Campaign | 0.520*** | 0.080 |
| | (0.075) | (0.094) |
| Subsidy & Camp & Conven | 0.458*** | 0.397*** |
| | (0.064) | (0.083) |
| R-squared | 0.318 | 0.166 |
| Mean | 0.504 | 0.379 |
| Control group mean | 0.272 | 0.230 |
| Number of observations | 4,168 | 3,415 |
| **P-value on test of equality** | | |
| Sub + Camp = Sub & Camp | 0.477 | 0.330 |
| Sub + Conv = Sub & Conv | 0.323 | 0.090 |
| Camp + Conv = Camp & Conv | 0.756 | 0.770 |
| Sub + Camp + Conv = Sub & Camp & Conv | 0.211 | 0.991 |

Notes: This table presents the effects of original intervention on enrollment in health insurance in short and long run. All regressions include a standard set of covariates (individual, household, and community) and baseline measure of dependent variable. *P*-values for the equality of effect estimates are also presented. Robust standard errors clustered at the community level are reported in parentheses. *, **, and *** denote statistical significance at 10 %, 5 %, and 1 % levels, respectively.

Table B.2: Effects on Enrollment with Restricted Sample.

| | Enrollment | |
|---|---|---|
| | Short-run | Long-run |
| | (1) | (4) |
| **Panel A** | | |
| Any Subsidy | 0.424*** | 0.135* |
| | (0.044) | (0.074) |
| R-squared | 0.399 | 0.228 |
| **Panel B** | | |
| Partial subsidy (positive price) | 0.405*** | 0.095 |
| | (0.047) | (0.067) |
| Full subsidy (free) | 0.514*** | 0.317*** |
| | (0.090) | (0.105) |
| R-squared | 0.401 | 0.238 |
| **Panel C** | | |
| 1/3 subsidy | 0.387*** | 0.137* |
| | (0.084) | (0.080) |
| 2/3 subsidy | 0.419*** | 0.063 |
| | (0.062) | (0.067) |
| Full subsidy (free) | 0.514*** | 0.316*** |
| | (0.090) | (0.105) |
| R-squared | 0.401 | 0.239 |
| Mean | 0.405 | 0.290 |
| Control group mean | 0.272 | 0.230 |
| Number of observations | 1,614 | 1,304 |

Notes: This table corresponds to Table 2, but the sample is restricted to subsidy only and control groups. All regressions include a standard set of covariates (individual, household, and community) and baseline measure of dependent variable. Robust standard errors clustered at community level are reported in parentheses. *, **, and *** denote statistical significance at 10 %, 5 %, and 1 % levels, respectively.

Table B.3: Effects on Health Care Utilization with Restricted Sample (Short Run).

| | Short run | | | | | |
|---|---|---|---|---|---|---|
| | Visited health facility in last four weeks | Visited health facility in last six months | # of visits in last four weekss | Visited Facility for malaria treatment in the last four weeks | Made an out-of-pocket for health service in the last six months | Standardized treatment effects |
| | (1) | (2) | (3) | (4) | (5) | (6) |
| **Panel A** | | | | | | |
| Any subsidy | -0.011 | -0.009 | -0.007 | 0.011 | -0.018 | -0.003 |
| | (0.010) | (0.020) | (0.024) | (0.008) | (0.016) | (0.009) |
| R-squared | 0.121 | 0.137 | 0.142 | 0.113 | 0.139 | 0.086 |
| **Panel B** | | | | | | |
| Partial subsidy | -0.012 | -0.005 | -0.004 | 0.014 | -0.011 | 0.001 |
| | (0.012) | (0.021) | (0.028) | (0.010) | (0.017) | (0.011) |
| Full subsidy | -0.001 | -0.028 | -0.025 | -0.010 | -0.049* | -0.020 |
| | (0.020) | (0.044) | (0.019) | (0.013) | (0.025) | (0.013) |
| R-squared | 0.121 | 0.137 | 0.143 | 0.114 | 0.141 | 0.086 |
| **Panel C** | | | | | | |
| 1/3 subsidy | -0.019 | -0.014 | -0.027 | 0.016 | -0.021 | -0.002 |
| | (0.020) | (0.023) | (0.038) | (0.012) | (0.016) | (0.013) |
| 2/3 subsidy | -0.006 | 0.001 | 0.014 | 0.013 | -0.003 | 0.003 |
| | (0.015) | (0.030) | (0.032) | (0.011) | (0.024) | (0.013) |
| Full subsidy | -0.001 | -0.028 | -0.025 | -0.010 | -0.049* | -0.020 |
| | (0.019) | (0.044) | (0.018) | (0.013) | (0.025) | (0.013) |
| R-squared | 0.121 | 0.138 | 0.144 | 0.114 | 0.141 | 0.086 |
| Control group mean | 0.038 | 0.101 | 0.033 | 0.018 | 0.046 | -0.011 |
| Number of observations | 1,200 | 1,566 | 1,196 | 1,263 | 1,622 | 6,191 |

Notes: This table corresponds to Table 5, but the sample is restricted to subsidy only and control groups. All regressions include a standard set of covariates (individual, household, and community) and baseline measure of dependent variable. Robust standard errors clustered at community level are reported in parentheses. *, **, and *** denote statistical significance at 10 %, 5 %, and 1 % levels, respectively.

Table B.4: Effects on Health Care Utilization with Restricted Sample (Long Run).

| | Long run | | | | | |
|---|---|---|---|---|---|---|
| | Visited health facility in last four weeks | Visited health facility in last six months | # of visits in last four weeks | Visited Facility for malaria treatment in the last four weeks | Made an out-of-pocket for health service in the last six months | Standardized treatment effects |
| | (1) | (2) | (3) | (4) | (5) | (6) |
| **Panel A** | | | | | | |
| Any subsidy | 0.041*** | 0.096*** | 0.031** | 0.028* | 0.003 | 0.039*** |
| | (0.013) | (0.026) | (0.012) | (0.014) | (0.006) | (0.010) |
| R-squared | 0.109 | 0.121 | 0.106 | 0.104 | 0.090 | 0.092 |
| **Panel B** | | | | | | |
| Partial subsidy | 0.045*** | 0.086*** | 0.033** | 0.030** | 0.005 | 0.041*** |
| | (0.013) | (0.020) | (0.012) | (0.014) | (0.007) | (0.010) |
| Full subsidy | 0.020 | 0.146** | 0.024 | 0.020 | -0.006 | 0.029 |
| | (0.018) | (0.060) | (0.020) | (0.019) | (0.009) | (0.019) |
| R-squared | 0.110 | 0.123 | 0.106 | 0.104 | 0.091 | 0.092 |
| **Panel C** | | | | | | |
| 1/3 subsidy | 0.012 | 0.071*** | 0.012 | 0.013 | -0.001 | 0.025** |
| | (0.010) | (0.025) | (0.009) | (0.009) | (0.009) | (0.007) |
| 2/3 subsidy | 0.070*** | 0.097*** | 0.048** | 0.043* | 0.009 | 0.053*** |
| | (0.018) | (0.032) | (0.019) | (0.021) | (0.009) | (0.017) |
| Full subsidy | 0.020 | 0.146** | 0.024 | 0.020 | -0.006 | 0.030 |
| | (0.019) | (0.060) | (0.020) | (0.020) | (0.009) | (0.019) |
| R-squared | 0.117 | 0.123 | 0.110 | 0.107 | 0.091 | 0.094 |
| Control group mean | 0.014 | 0.044 | 0.011 | 0.009 | 0.012 | -0.021 |
| Number of observations | 1,236 | 1,546 | 1,238 | 1,236 | 1,546 | 6,180 |

Notes: This table corresponds to Table 6, but the sample is restricted to subsidy only and control groups. All regressions include a standard set of covariates (individual, household, and community) and baseline measure of dependent variable. Robust standard errors clustered at community level are reported in parentheses. *, **, and *** denote statistical significance at 10 %, 5 %, and 1 % levels, respectively.

Table B.5: Effects on Health Status with Restricted Sample (Short Run).

| | Short run | | | | |
|---|---|---|---|---|---|
| | Healthy or very healthy | # Days ill last four weeks | Could not perform normal daily activities due to illness last four weeks | # days could not perform normal daily activities due to illness in the last four weeks | Standardized treatment effects |
| | (1) | (2) | (3) | (4) | (5) |
| **Panel A** | | | | | |
| Any subsidy | 0.148*** | -0.421** | -0.025 | -0.315 | -0.037** |
| | (0.043) | (0.185) | (0.017) | (0.447) | (0.017) |
| R-squared | 0.346 | 0.139 | 0.136 | 0.141 | 0.107 |
| **Panel B** | | | | | |
| Partial subsidy (positive price) | 0.152*** | -0.445** | -0.021 | -0.203 | -0.033* |
| | (0.044) | (0.196) | (0.019) | (0.497) | (0.018) |
| Full subsidy (free) | 0.130* | -0.308 | -0.041 | -0.854 | -0.058** |
| | (0.076) | (0.283) | (0.038) | (0.619) | (0.027) |
| R-squared | 0.346 | 0.139 | 0.136 | 0.142 | 0.107 |
| **Panel C** | | | | | |
| 1/3 subsidy | 0.157*** | -0.740** | -0.044 | -0.916 | -0.061** |
| | (0.047) | (0.300) | (0.031) | (0.728) | (0.027) |
| 2/3 subsidy | 0.147** | -0.225 | -0.005 | 0.343 | -0.011 |
| | (0.061) | (0.250) | (0.021) | (0.517) | (0.021) |
| Full subsidy (free) | 0.130 | -0.298 | -0.040 | -0.836 | -0.057** |
| | (0.077) | (0.287) | (0.038) | (0.617) | (0.027) |
| R-squared | 0.346 | 0.141 | 0.137 | 0.145 | 0.109 |
| Control group mean | 0.818 | 0.616 | 0.082 | 1.376 | 0.011 |
| Number of observations | 478 | 1,597 | 1,603 | 1,549 | 5,081 |

Notes: This table corresponds to Table 7, but the sample is restricted to subsidy only and control groups. All regressions include a standard set of covariates (individual, household, and community) and baseline measure of dependent variable. Robust standard errors clustered at community level are reported in parentheses. *, **, and *** denote statistical significance at 10 %, 5 %, and 1 % levels, respectively.

Table B.6: Effects on Health Status with Restricted Sample (Long Run).

| | Long run | | | | |
|---|---|---|---|---|---|
| | Healthy or very healthy | # Days ill last four weeks | Could not perform normal daily activities due to illness last four weeks | # days could not perform normal daily activities due to illness in the last four weeks | Standardized treatment effects |
| | (1) | (2) | (3) | (4) | (5) |
| **Panel A** | | | | | |
| Any subsidy | -0.128** | 0.249** | 0.042*** | 0.243** | 0.050*** |
| | (0.047) | (0.097) | (0.013) | (0.090) | (0.014) |
| R-squared | 0.416 | 0.095 | 0.132 | 0.094 | 0.083 |
| **Panel B** | | | | | |
| Partial subsidy | -0.133** | 0.296*** | 0.045*** | 0.268** | 0.055*** |
| | (0.058) | (0.095) | (0.013) | (0.097) | (0.014) |
| Full subsidy | -0.111 | 0.019 | 0.027* | 0.122 | 0.026 |
| | (0.126) | (0.157) | (0.013) | (0.097) | (0.020) |
| R-squared | 0.416 | 0.096 | 0.133 | 0.095 | 0.084 |
| **Panel C** | | | | | |
| 1/3 subsidy | -0.115** | 0.287** | 0.027*** | 0.253** | 0.045*** |
| | (0.051) | (0.107) | (0.010) | (0.109) | (0.014) |
| 2/3 subsidy | -0.149 | 0.303** | 0.059*** | 0.279** | 0.063*** |
| | (0.088) | (0.128) | (0.018) | (0.123) | (0.019) |
| Full subsidy | -0.110 | 0.019 | 0.027* | 0.123 | 0.026 |
| | (0.127) | (0.157) | (0.013) | (0.098) | (0.020) |
| R-squared | 0.416 | 0.096 | 0.135 | 0.095 | 0.084 |
| Control group mean | 0.792 | 0.355 | 0.012 | 0.083 | -0.019 |
| Number of observations | 416 | 1,531 | 1,530 | 1,475 | 4,814 |

Notes: This table corresponds to Table 8, but the sample is restricted to subsidy only and control groups. All regressions include a standard set of covariates (individual, household, and community) and baseline measure of dependent variable. Robust standard errors clustered at community level are reported in parentheses. *, **, and *** denote statistical significance at 10 %, 5 %, and 1 % levels, respectively.